\documentclass[aps,prx,reprint,superscriptaddress,amsmath,amssymb,a4paper,twocolumn,longbibliography]{revtex4-2}
\usepackage{amsmath}
\usepackage{amsfonts}
\usepackage{amssymb}
\usepackage{graphicx}
\usepackage{longtable}
\usepackage{dcolumn}
\usepackage{bm}
\usepackage{epstopdf}
\usepackage{datetime}
\usepackage{comment}
\usepackage{siunitx}
\DeclareSIUnit\bit{bit}
\usepackage{physics}
\usepackage{color}

\graphicspath{{.}{Figures/}}

\makeatletter
\def\maketitle{
\@author@finish
\title@column\titleblock@produce
\suppressfloats[t]}
\makeatother

\begin{document}
\title{Nonreciprocal buckling makes active filaments polyfunctional}

\author{Sami C.~Al-Izzi}
\thanks{These authors contributed equally}
\affiliation{School of Physics, UNSW, Sydney, NSW 2052, Australia.}
\affiliation{ARC Centre of Excellence for the Mathematical Analysis of Cellular Systems - UNSW Node, Sydney, NSW 2052, Australia.}
\affiliation{Department of Mathematics, Faculty of Mathematics and Natural Sciences, University of Oslo, 0315 Oslo, Norway}

\author{Yao Du}
\thanks{These authors contributed equally}
\affiliation{Institute of Physics, Universiteit van Amsterdam, Science Park 904, 1098 XH Amsterdam, The Netherlands}

\author{Jonas Veenstra}
\affiliation{Institute of Physics, Universiteit van Amsterdam, Science Park 904, 1098 XH 
Amsterdam, The Netherlands}

\author{Richard G.~Morris}
\affiliation{School of Physics, UNSW, Sydney, NSW 2052, Australia.}
\affiliation{ARC Centre of Excellence for the Mathematical Analysis of Cellular Systems - UNSW Node, Sydney, NSW 2052, Australia.}

\author{Anton Souslov}
\affiliation{TCM Group, Cavendish Laboratory, University of Cambridge, Cambridge CB3 0US, United Kingdom}

\author{Andreas Carlson}
\affiliation{Department of Mathematics, Faculty of Mathematics and Natural Sciences, University of Oslo, 0315 Oslo, Norway}

\author{Corentin Coulais}
\affiliation{Institute of Physics, Universiteit van Amsterdam, Science Park 904, 1098 XH Amsterdam, The Netherlands}

\author{Jack Binysh}
\email{j.a.c.binysh@uva.nl}
\affiliation{Institute of Physics, Universiteit van Amsterdam, Science Park 904, 1098 XH Amsterdam, The Netherlands}

\begin{abstract}
Active filaments are a workhorse for propulsion and actuation across biology, soft robotics and mechanical metamaterials. However, artificial active rods suffer from limited robustness and adaptivity because they rely on external control, or are tethered to a substrate. Here we bypass these constraints by demonstrating that non-reciprocal interactions lead to large-scale unidirectional dynamics in free-standing slender structures. By coupling the bending modes of a buckled beam anti-symmetrically, we transform the multistable dynamics of elastic snap-through into persistent cycles of shape change. In contrast to the critical point underpinning beam buckling, this transition to self-snapping is mediated by a critical exceptional point, at which bending modes simultaneously become unstable and degenerate. Upon environmental perturbation, our active filaments exploit self-snapping for a range of functionality including crawling, digging and walking. Our work advances critical exceptional physics as a guiding principle for programming instabilities into functional active materials.
\end{abstract}

\maketitle
\begin{figure*}[t]
    \centering
    \includegraphics{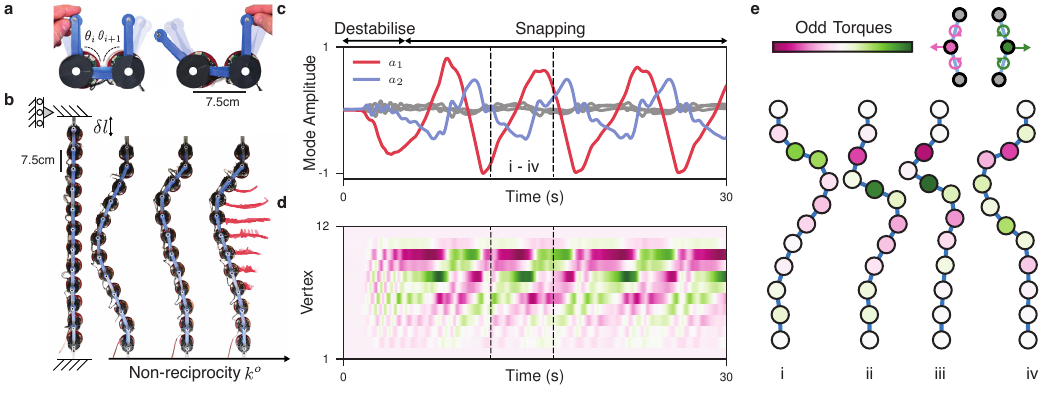}
    \caption{\textbf{Non-reciprocal Buckling.} Adding non-reciprocity to the canonical Euler buckling scenario causes beams to lose stability and transition into self-snapping. 
    \textbf{(a)} Within a building block, each linkage has a torque-angle relationship $ \tau_i =k^\mathrm{o}(\delta \theta_{i+1} -\delta \theta_{i-1})$, and responds antisymmetrically when perturbed from the left vs. right. 
    \textbf{(b)} Buckling a chain of non-reciprocal linkages under clamped tangent boundary conditions, we find that as non-reciprocity $k^\mathrm{o}$ increases, the beam polarizes but retains stability. Increasing activity still further, this stable state disappears and the beam persistently self-oscillates. Traces indicate linkage positions over time. Scale bar $7.5\ \mathrm{cm}$.
    \textbf{(c)} Decomposing the horizontal position of each linkage into normalized Fourier modes $a_k$, we find that the lowest two modes dominate the snapping process, oscillating out of phase with one another. Modes $a_k$ for $k>2$ are shown in gray, and remain negligible during snapping. 
    \textbf{(d)} Smooth waves of odd torques run down the filament during snapping. 
    \textbf{(e)} A single snap proceeds via a pulsation traveling from the tail of the filament to its head. The schematic indicates the torque dipoles exerted by each motor, alongside the force on the central linkage.}
    \label{fig:buckling}
\end{figure*}

Take a paper ticket and compress it between two of your fingers. It will spontaneously lose its stability and buckle one way or the other. Now try to push the buckled state inwards with your other hand. It will first resist, but then suddenly snap to the other side. Such buckling and snapping is a quintessential example of a critical bifurcation and has many uses: from carnivorous plants and beetles~\cite{
wangInsectscaleJumpingRobots2023,
forterreHowVenusFlytrap2005} to micro-electromechanical systems~\cite{brennerOptimalDesignBistable2003}, soft robots~\cite{xiEmergentBehaviorsBucklingdriven2024}
and mechanical metamaterials~\cite{dengNonlinearWavesFlexible2021,dykstraBucklingMetamaterialsExtreme2023,siefertBioinspiredPneumaticShapemorphing2019,
dudekShapemorphingMetamaterials2025}, 
buckling endows slender objects with sensitivity to environmental perturbations, multistable response and the capacity for large-scale shape morphing. Buckling and snapping typically occur once---they only happen if you push the beam. If instead one wishes to obtain persistent oscillations to e.g., perform work via cyclic shape changes, then one would need to constantly energize the beam from within. 

Such design principles are evolved and exploited in living systems such as flagella, worms, and snakes~\cite{cammannFormFunctionBiological2025a,huMechanicsSlitheringLocomotion2009, guoLimblessUndulatoryPropulsion2008,patilUltrafastReversibleSelfassembly2023,cassReactiondiffusionBasisAnimated2023}. They all consume energy, spontaneously compress and bend, and undergo autonomous cycles of shape changes which power swimming, pumping, crawling, and burrowing. Some of these biological filaments even combine activity with elastic buckling and snapping, which allows them to perform fast actions such as reorientation~\cite{wangMechanicalIntelligenceSimplifies2023, sonBacteriaCanExploit2013}. Surprisingly, the precise interplay between activity, buckling and snapping remains poorly understood. We lack model systems to precisely capture the dynamics in the vicinity of those bifurcations.

Here we show that active beams with non-reciprocal internal mechanics buckle via a critical exceptional point (CEP)~\cite{braunsNonreciprocalPatternFormation2024, suchanekEntropyProductionNonreciprocal2023, fruchartNonreciprocalPhaseTransitions2021, hanaiCriticalFluctuationsManybody2020}. This scenario is characterized by two ingredients: the mechanical criticality of buckling plus non-Hermitian linear response~\cite{chenRealizationActiveMetamaterials2021, chengOddElasticityRealized2021, fossatiOddElasticityTopological2024, nemethNonreciprocalConstitutiveLaws2025, jiangEnergyPrinciplesNonHermitian2025}. In conventional buckling, a single mode of a stressed beam becomes unstable via a critical point. By contrast, in CEP-mediated buckling two anti-symmetrically coupled modes of an active beam simultaneously destabilize. The resulting run-and-chase dynamics between these modes produces large oscillations in beam shape. We experimentally demonstrate that these stable oscillations are limit cycles, and develop a theoretical model that captures the nonlinear dynamics of the beam. These limit cycles are distinct from a transition to linear oscillations across exceptional points~\cite{miriExceptionalPointsOptics2019}, because these cycles arise from the supercritical instability and post-bifurcation nonlinear dynamics of beam buckling. Finally, we harness these dynamics for multiple robotic functionalities---we demonstrate crawling, walking and digging in a single free-standing filament, without the need for external forcing~\cite{sekimotoSymmetryBreakingInstabilities1995,decanioSpontaneousOscillationsElastic2017, filyBucklingInstabilitiesSpatiotemporal2020,clarkeBifurcationsNonlinearDynamics2024,zhengSelfOscillationSynchronizationTransitions2023,sinaasappelParticleSweepingCollection2026,weiAutonomousLifelikeBehavior2025}. 
Our beams are made by chaining together mechatronic linkages [Fig.~\ref{fig:buckling}(a), Methods~\S 4]. At the level of a single linkage, conventional bending elasticity is encoded in a torque-angle relation $\tau_i =-B\delta \theta_i$ at site $i$. In contrast, our linkages implement a torque-angle relation that couples neighboring angles antisymmetrically: $\tau_i =k^\mathrm{o}(\delta \theta_{i+1} -\delta \theta_{i-1})$, where $k^\mathrm{o}$ quantifies the microscopic non-reciprocity. This antisymmetric relation cannot be derived from a potential energy~\cite{veenstraAdaptiveLocomotionActive2025}, and injects energy into the beam in a manner that breaks parity symmetry whilst conserving linear and angular momentum. 
Whilst energy is injected into the beam by non-reciprocal couplings, it is dissipated via damping at the linkage hinges. We control dissipation by attaching a viscoelastic polymer skeleton to the backbone of the linkage. This skeleton provides bending rigidity $B$ but also acts as a torsional dashpot on each hinge, $\tau_i= - \Gamma \delta \dot{{\theta}}_i$, with a damping $\Gamma$---we calibrate these values by fitting to the dynamics of a damped torsional oscillator [Methods \S\ref{sec:Protocol}, Figs.~\ref{fig:Agilus_kappa}, \ref{fig:Agilus_Gamma}].

We buckle an open segment of our active beam under clamped tangent boundary conditions with a compression $\mathrm{d} l$ [Fig.~\ref{fig:buckling}(b), Methods Fig.~\ref{fig:Setup_buckle}, Supplementary Video 1]. As we increase the activity $k^\mathrm{o}$, the initially symmetric buckled state polarizes, leaning to one side but remaining static.
However, above a critical $k^\mathrm{o}$, this static shape loses stability and a persistent self-snapping sets in. 
Without a source of damping within the hinges, many modes participate in this snapping dynamics. Yet by increasing dissipation via our viscoelastic skeleton [using Stratasys Agilus 30, calibration in Figs.~\ref{fig:Agilus_kappa},\ref{fig:Agilus_Gamma}], we drive the chain into a long-wavelength regime, where snapping is dominated only by the lowest Fourier modes [Fig.~\ref{fig:buckling}(c)].
Visualizing the non-reciprocal torque distribution, we find smooth stress waves that pulse down the chain in a direction determined by $k^\mathrm{o}$, snapping the filament back and forth [Fig.~\ref{fig:buckling}(d, e)]. Upon removing compression, snapping ceases and the beam stabilizes. We conclude that the nonlinear oscillations we observe are being driven by a combination of two phenomena: polarization induced by non-reciprocity and buckling induced by confinement.

\begin{figure*}[t]
    \includegraphics{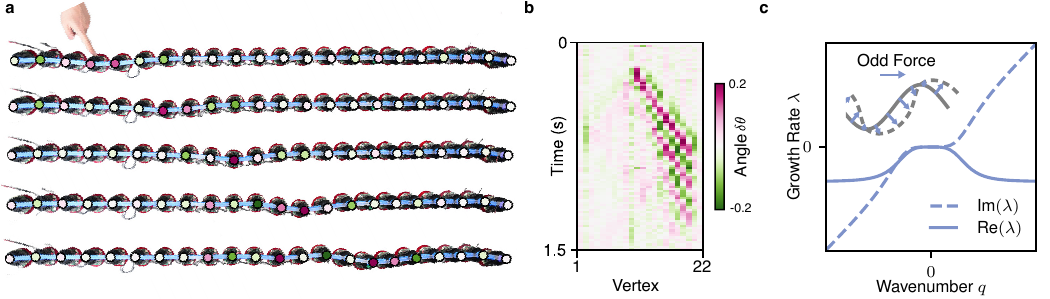}
    \caption{ {\bf Breaking reciprocity in slender filaments creates one-way flexural waves}. 
    \textbf{(a, b)} Perturbing an uncompressed, loosely confined chain of non-reciprocal linkages, we observe one-way advection of curvature. Panel (a) shows key frames of a perturbation applied at one free end. Panel (b) shows a perturbation applied at the chain midpoint, confirming the asymmetry of wave propagation. Colors indicate angular deviations from the flat state.  
    \textbf{(c)} Non-reciprocal forces on a sinusoidal filament show a $\pi/2$ phase lag with respect to the filament geometry itself, leading to an antisymmetric flexural wave dispersion relation. Here $\lambda(q)$ is the complex dispersion from~\eqref{eq:RingDispersion}, with $\mathrm{Im}(\lambda)$ representing wave propagation and $\mathrm{Re}(\lambda)$ giving growth or decay. Dispersion shown for $\beta=\eta=1$.}
    \label{fig:Mechanism}
\end{figure*}

First, we focus on the polarization of the beam by removing external compression and studying the linear response of a stable filament. 
We loosely confine a beam under zero compression, and apply a perturbation to its left edge [Fig.~\ref{fig:Mechanism}(a), Methods~\S 3]. In a passive beam, waves would propagate and damp equally in both directions. Yet here we observe that the non-reciprocity $k^\mathrm{o}$ picks a clear direction of wave motion along the filament [Fig.~\ref{fig:Mechanism}(b), Supplementary Video 2]: The polarization of our clamped beam is being driven by one-way wave advection of curvature.

To rationalize this advective dynamics, we revisit the continuum mechanics of a slender beam. A passive beam under line tension $\Lambda$ and bending rigidity $A$ experiences a force ${\bf f} \sim \Lambda \kappa {\bf n} + A \partial_s^{2}\kappa {\bf n} + \hdots$ 
, where $\kappa$ is the geometric curvature of the beam and ${\bf n}$ is its normal. These forces only depend on even gradients of $\kappa$, a restriction that is enforced by energy conservation. The simplest non-conservative modification to the mechanics of the beam is to introduce internal stresses that break this parity symmetry, which to lowest order in curvature read
\begin{equation}
    {\bf f} =\zeta \partial_s^{3}\kappa {\bf n},
    \label{eq:EoM}
\end{equation}
where $\zeta$ is the non-reciprocity. We now ask: How do undulations propagate in the presence of $\zeta$?

We compute ${\bf f}$ for a sinusoidal filament profile ${\bf x}(s)=\sin(q s)$ [Fig.~\ref{fig:Mechanism}(c)]. For passive elastic response ${\bf f} \sim Aq^4 \sin(q s)$: The forcing is in phase with the filament geometry and flattens variations in curvature. By contrast, in the purely non-reciprocal limit
${\bf f} \sim \zeta q^{5} \cos(q s)$, a forcing that is $\pi/2$ out of phase with the filament geometry. This lagged response is ubiquitous across non-reciprocal spin~\cite{hanaiNonreciprocalFrustrationTime2024} and soft-matter systems~\cite{braunsNonreciprocalPatternFormation2024} and advects patterns without attenuation. 

At linear order, and in the limit of large damping, the forcing term~\eqref{eq:EoM} modifies the dynamics of a close-to-flat filament with deflection $h(x,t)$ to give the following non-reciprocal dynamic beam equation [Methods \S\ref{sec:Continuum}]:
\begin{equation}
    \beta \dot{h} =  (\Lambda \partial_x^2 - A \partial_x^4  - \zeta \partial_x^5 - \eta\partial_x^4\partial_t)h\text{,}
    \label{eq:NRBeam}
\end{equation}
where $\beta$ describes substrate friction and $\eta$ describes viscous dissipation by beam bending. Computing the dispersion relation of~\eqref{eq:NRBeam} we find that each Fourier mode $q$ has a growth rate $\lambda(q)$ of
\begin{align}
    \lambda(q) = \frac{-1}{\beta+\eta q^4} \left(\Lambda q^2 + A q^4 + \mathrm{i}\zeta q^5\right)\text{.}
    \label{eq:RingDispersion}
\end{align}
The non-reciprocity $\zeta$ introduces a complex component to $\lambda(q)$, causing advection even in highly viscous environments. We now use the dispersion relation~\eqref{eq:RingDispersion} to construct a scaling argument for the onset of snapping under confinement.

When a beam of length $L$ is confined, Euler buckling can be predicted by balancing line tension against bending elasticity to give $\Lambda/A\sim1/L^2$.Equation~\eqref{eq:RingDispersion} suggests that the onset of snapping from within the buckled state comes from a second balance, namely between restoring elasticity and non-reciprocity: $ \zeta/A \sim L$. To connect this continuum expression to our experimental parameters, we note that $\zeta$ is analogous to the bending modulus $A$ but coupled to a single extra gradient. The microscopic passive stiffness $B$ coarse-grains to the continuum $A$ as $A \sim B l_0 $, with $l_0$ the length of a single linkage. By analogy, $k^\mathrm{o}$---which responds to microscopic angular gradients---coarse-grains to $\zeta$ as $\zeta \sim k^\mathrm{o} l_0^2$. Expressing this second balance with our microscopic parameters, we hypothesize that the onset of snapping occurs when $k^\mathrm{o}/B \sim 1$.  

\begin{figure*}[t!]
    \centering
    \includegraphics{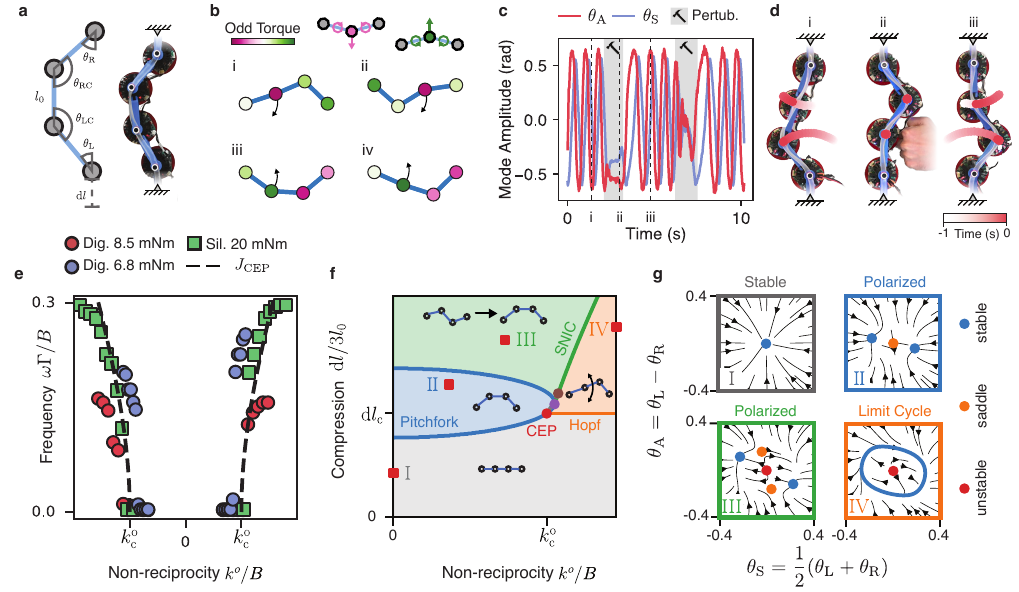}
    \caption{
    \textbf{The odd von Mises truss is a minimal model of non-reciprocal buckling.} We capture non-reciprocal buckling in an odd von Mises truss with clamped ends. We compress this truss a distance $\text{d}l$ from its rest length. As $k^\mathrm{o}$ increases, the truss polarizes, and then begins to self-snap. 
    \textbf{(a)} Schematic and experimental realization of the truss, with internal angles  ($\theta_\mathrm{L},\theta_\mathrm{LC},\theta_\mathrm{RC},\theta_\mathrm{R}$) and linkage rest length $l_0$.
    \textbf{(b)} The non-reciprocal torques at each vertex during a single snap-through. 
    \textbf{(c, d)} Plotting the dynamics of the truss in terms of its symmetric $\theta_{\text{s}} = (\theta_L + \theta_R)/2$ and  antisymmetric $\theta_\text{A} = (\theta_L - \theta_R)$ modes, we find that self-oscillations reliably return to the same amplitude and frequency after multiple environmental perturbations, an experimental signature of an attractive limit cycle.
    \textbf{(e)} We measure the  self-snapping frequency $\omega$ as a function of $k^\mathrm{o}$ for a range of passive bending stiffness $B$: $B=8.5\text{ mNm}$ and $B=6.8\text{ mNm}$ (circles) are implemented via electronic feedback, with $B=20\text{ mNm}$ (squares) implemented via a silicone elastomer. Rescaling the activity by this bending rigidity, $k^\mathrm{o}/B$, and the snapping frequency $\omega$ by the characteristic dissipative timescale $\omega \Gamma /B$ collapses these data onto a single curve given by Eq.~\ref{eq:CEP}. We find a square-root growth of frequencies $\omega\sim \sqrt{k^\mathrm{o}-k^\mathrm{o}_c}$ beyond a critical activity $k^\mathrm{o}_c$ that is characteristic of an exceptional transition. 
    \textbf{(f, g)} Our theory Eq.~\ref{eq:TrussDynamics}  gives a complete phase diagram of the bifurcations in this minimal system.  This shows a stable region (gray) and buckling phases: at lower compression a simple buckling (blue), then at higher compressions buckling with transient jackknifing (green), along with the snapping phase where the non-reciprocity creates a limit cycle in phase space (orange). Crucially these four regions encircle a critical exceptional point, CEP, red dot in panel (f). The purple and brown dots correspond to a cusp bifurcation and a global bifurcation where the narrow region of multi-stable buckling and snapping collapses to a global SNIC bifurcation.
    }
    \label{fig:Truss}
\end{figure*}

Motivated by the observation that snapping is dominated by the lowest modes of the beam, to capture the essence of the transition we therefore construct a chain with only two angular degrees of freedom---an odd von Mises truss [Fig.~\ref{fig:Truss}(a), Supplementary Video 3]~\cite{wangTransientAmplificationBroken2024}. Compressing the truss and increasing $k^\mathrm{o}$, we find the same polarization and transition to self-snapping observed in our longer chains [Fig.~\ref{fig:Truss}(b)]. 
At a given $k^\mathrm{o}$, these self-oscillations reliably converge to the same amplitude and frequency for different initial conditions [Figs.~\ref{fig:Truss}(c,d), Supplementary Video 5]. This robustness is a strong experimental signature of a limit cycle---a stable attractor for the nonlinear dynamics of the truss.

We now track the snapping frequency $\omega$ as $k^\mathrm{o}$ is increased, for several values of the passive stiffness $B$ and dissipation $\Gamma$---we tune these parameters by varying both the viscoelastic skeleton and through additional motor torques that digitally implement $B$ [Methods \S\ref{sec:Protocol}].
When we rescale our $k^\mathrm{o}$ values by the passive stiffness $B$ as $k^\mathrm{o}/B$, and our snapping frequencies $\omega$ by the viscous timescale $\Gamma/B$ as $\omega \Gamma /B$, we 
 collapse data for multiple stiffnesses onto a single critical threshold $k^\mathrm{o}_c$ in $k^\mathrm{o}/B$ [Fig.~\ref{fig:Truss}(e)]. However, we also find that oscillations do not emerge at a finite frequency, as per a standard Hopf bifurcation~\cite{zhengSelfOscillationSynchronizationTransitions2023}. Instead, their frequency grows as $\omega \sim \sqrt{k^\mathrm{o} - k^\mathrm{o}_\mathrm{c}}$, i.e. a square-root singularity that is the hallmark of an exceptional transition~\cite{miriExceptionalPointsOptics2019}. We therefore hypothesize that the nonlinear limit-cycle dynamics appear as the beam crosses an exceptional transition in its linear response.
 
 To explore this bifurcation scenario we develop a detailed model of the truss dynamics. Each node in the truss experiences a non-reciprocal angular bond tension $\tau^\text{o}_i=k^\mathrm{o}(\delta \theta_{i+1} - \delta \theta_{i-1})$, passive bending elasticity $\tau^\text{el}_i=-B\delta\theta_i$, and viscous dissipation $\tau^\text{vis}_i=-\Gamma \delta \dot{\theta}_i$. We additionally allow for a finite longitudinal spring stiffness $k$, which is typically large, $k\gg B l_0^2$ for a linkage spacing $l_0$. The mechanical energy of this system is given by
 \begin{equation}
     E=\sum_i \frac{B}{2}\delta\theta_i^2 + \sum_e\frac{k}{2}\left(|l_e|-l_0\right)^2\text{,}
     \label{eq:PassiveMain}
 \end{equation}
 where $|l_e|$ is the current length of longitudinal spring $e$. The summation over $i$ in~\eqref{eq:PassiveMain} refers to nodes, and over $e$ to  connecting longitudinal springs. The dissipative forces and active torques can be combined into a Rayleigh dissipation function using the principle of virtual work,
 \begin{equation}
     \mathcal{R}=\sum_i\left(\frac{\Gamma}{2}\delta \dot{\theta}_i^2 + \tau^\text{o}_i\delta \dot{\theta}_i\right)\text{.}
     \label{eq:DissipativeMain}
 \end{equation}

We now consider the limit of large longitudinal spring constant $k$, in which the spring lengths equilibrate much faster than the torsional springs. We then derive dynamical equations for the angular degrees of freedom in the limit of small deflections using the variational principle
\begin{equation}
    \frac{\partial E}{\partial \theta_i }+ \frac{\partial \mathcal{R}}{\partial \dot\theta_i} = 0\mathrm{.}
    \label{eq:RayleighMain}
\end{equation}
Next, we expand~\eqref{eq:RayleighMain} to cubic order in powers of the strain $\delta l/l$, where $\delta l$ is the compression of the truss and $l=3l_0$ is the uncompressed truss rest length [Fig.~\ref{fig:Truss}(a)]---these algebraic manipulations are given in Methods \S\ref{sec:TrussTheory}.
The resulting dynamics,~\eqref{eq:RayleighMain}, is conveniently expressed in terms of the symmetric $\theta_\text{S} = \frac{1}{2}(\theta_\text{L}+\theta_\text{R})\text{,}$ and anti-symmetric $\theta_\text{A} = (\theta_\text{L}-\theta_\text{R})$ combinations of the end node angles $\theta_{L,R}$:
\begin{align}
\label{eq:TrussDynamics}
    \dot{{\boldsymbol{\theta}}}= J {\boldsymbol{\theta}} + 
    \sum_{j=0}^{3}
         {\bf A}_{m} \theta_\text{S}^m \theta_\text{A}^{3-m} \text{,}
\end{align}
where $\boldsymbol \theta = (\theta_S, \theta_A)$ and the coefficients ${\bf A}_{m}$ are functions of $k$, $k^\mathrm{o}$ and the compression $\mathrm{d}l$ as detailed in Methods \S \ref{sec:TrussTheory}. The up/down symmetry of buckling imposes a $\mathbb{Z}_2$ equivariance $(\theta_S, \theta_A) \rightarrow (-\theta_S, -\theta_A)$ on the dynamics ~\eqref{eq:TrussDynamics}, forcing odd powers of $\theta$ only, in analogy to a classic pitchfork bifurcation. The additional symmetry under simultaneous reflection and a reversal in the sense of non-reciprocity, $(\theta_S, \theta_A, k^\mathrm{o} )\rightarrow (\theta_S, -\theta_A, -k^\mathrm{o})$, constrains the functional forms of ${\bf A}_{m}$ and forces $k^\mathrm{o}$ to appear only in the off-diagonal elements of the linearized response matrix $J$. 

The bifurcation structure of~\eqref{eq:TrussDynamics} is shown in Fig.~\ref{fig:Truss}(f). We find a line of exceptional transitions, which meet a second line of pitchfork/Hopf bifurcations at a single point, $(k_\mathrm{c}^\mathrm{o}, \mathrm{d}l_\mathrm{c}) = ({1}/{2 \sqrt{5}} +{108}/{\sqrt{5} k}, {45}/{2 k})$ in the limit of large longitudinal stiffness $k$. At this point, we have a double-zero eigenvalue in $J$: the combination of the critical phenomena of buckling and the non-Hermitian physics of non-reciprocity meets at a critical exceptional point (CEP)~\cite{hanaiCriticalFluctuationsManybody2020}. 
Expanding $J$ about the CEP as $k^\mathrm{o}= k^\mathrm{o}_\mathrm{c} + \delta k^\mathrm{o}$ and $\mathrm{d}l = \mathrm{d}l_\mathrm{c} + \delta l/k$ and taking the rigid linkage limit $k\to\infty$ we find
\begin{equation}
    {J}_{\text{CEP}} = \left(
\begin{array}{cc}
 \frac{1}{36} (2 \delta l  +9) & \frac{1}{8} \left(10 \delta k^\mathrm{o}+\sqrt{5}\right) \\
 -\delta k^\mathrm{o}-\frac{1}{2 \sqrt{5}} & \frac{\delta l }{30}-\frac{1}{4} \\
\end{array}
\right)\text{,}
\label{eq:CEP}
\end{equation}
which has eigenvalues $\lambda \sim \delta l + \sqrt{\delta l +  \delta k^\mathrm{o}}$, in excellent agreement with our experimentally obtained snapping frequencies [Fig.~\ref{fig:Truss}(e)].  

Our experimental cycle---first compression, next increasing activity, and finally relaxation---traces a loop that encircles the CEP in Fig.~\ref{fig:Truss}(f). Along this loop~\eqref{eq:TrussDynamics} predicts four phases, labeled (I)---(IV) in Figs.~\ref{fig:Truss}(f,g), which collectively form a $\mathbb{Z}_2$-symmetric Bogdanov-Takens bifurcation~\cite{golubitskyBurstingCoupledCell2005}. Here, the $\mathbb{Z}_2$ symmetry stems from the physics of buckling, but such discrete symmetries are prevalent in many minimal condensed matter systems~\cite{avniDynamicalPhaseTransitions2025}. (I) labels the flat unbuckled state [Figs.~\ref{fig:Truss}(f,g) gray region]. Starting from the flat state $(k^\mathrm{o}, \mathrm{d} l)=(0,0)$, upon increasing $\mathrm{d} l$ two symmetrically placed stable fixed points emerge as $\theta_S \sim \pm \sqrt{\delta l}$ in a pitchfork bifurcation, leaving the flat state as a marginally stable saddle point. Initially $\theta_A=0$ because the buckled state is left-right symmetric. As non-reciprocity $k^\mathrm{o}$ increases, the truss polarizes, and these stable points acquire nonzero $\theta_A$. The region with two stable buckled states is labeled (II) [Figs.~\ref{fig:Truss}(f,g) blue]. For sufficiently large compressions $\mathrm{d}l$, a new pair of saddle fixed points emerge from the flat state, representing transient jacknifed truss configurations. The flat state is now fully unstable. This region we denote (III), see Figs.~\ref{fig:Truss}(f, g), green. Upon further increase of $k^\mathrm{o}$ we cross the exceptional transition. At the nonlinear level, this transition to snapping manifests as a global saddle-node invariant circle (SNIC) bifurcation, in which each pair of stable/unstable fixed points merge to leave behind a single limit cycle. This snapping phase we denote (IV), see Figs.~\ref{fig:Truss}(f,g), orange. Finally, this limit cycle vanishes in a supercritical Hopf bifurcation as compression $\mathrm{d} l$ is decreased.

Our theory~\eqref{eq:TrussDynamics} predicts that close to the CEP, the SNIC bifurcation is in fact composed of several more intricate bifurcations, including a region of dynamical multistability, in which snapping and fixed buckled states exist simultaneously. In SI~\S 3 we characterize this region of global bifurcations precisely. For the truss, such multistability exists in a narrow region of parameter space; the purple and brown dots in Fig.~\ref{fig:Truss}(f) denote the cusp bifurcation where region (II) disappears and the point where the global multistable region collapses into a SNIC respectively.

The CEP-mediated bifurcation scenario we show in Fig.~\ref{fig:Truss} is ubiquitous across non-Hermitian critical systems. Analogous bifurcations have recently been proposed to give run-and-chase dynamics in  non-reciprocal spin~\cite{ hanaiCriticalFluctuationsManybody2020,zellaUniversal2024, hanaiNonreciprocalFrustrationTime2024, avniDynamicalPhaseTransitions2025} and reaction-diffusion~\cite{fruchartNonreciprocalPhaseTransitions2021,braunsNonreciprocalPatternFormation2024, suchanekEntropyProductionNonreciprocal2023} systems. Here the CEP exists within the modal structure of our beam: each species is a separate bending mode, and non-reciprocal couplings mix these usually orthogonal motions. This bifurcation scenario also applies to active beams more generally: in SI~\S 5 we show a CEP in the popular case of a buckled filament driven by tangential forces~\cite{sekimotoSymmetryBreakingInstabilities1995,decanioSpontaneousOscillationsElastic2017, filyBucklingInstabilitiesSpatiotemporal2020,clarkeBifurcationsNonlinearDynamics2024,zhengSelfOscillationSynchronizationTransitions2023,active_poly_review2}. The key ingredients are the criticality of buckling and the non-Hermitian mechanics induced by energy injection. Yet unlike tangential forces, which require a substrate, our non-reciprocal beams are free-standing, a feature we now exploit for robotic functions.

\begin{figure*}[t]
    \centering
    \includegraphics[width=\textwidth]{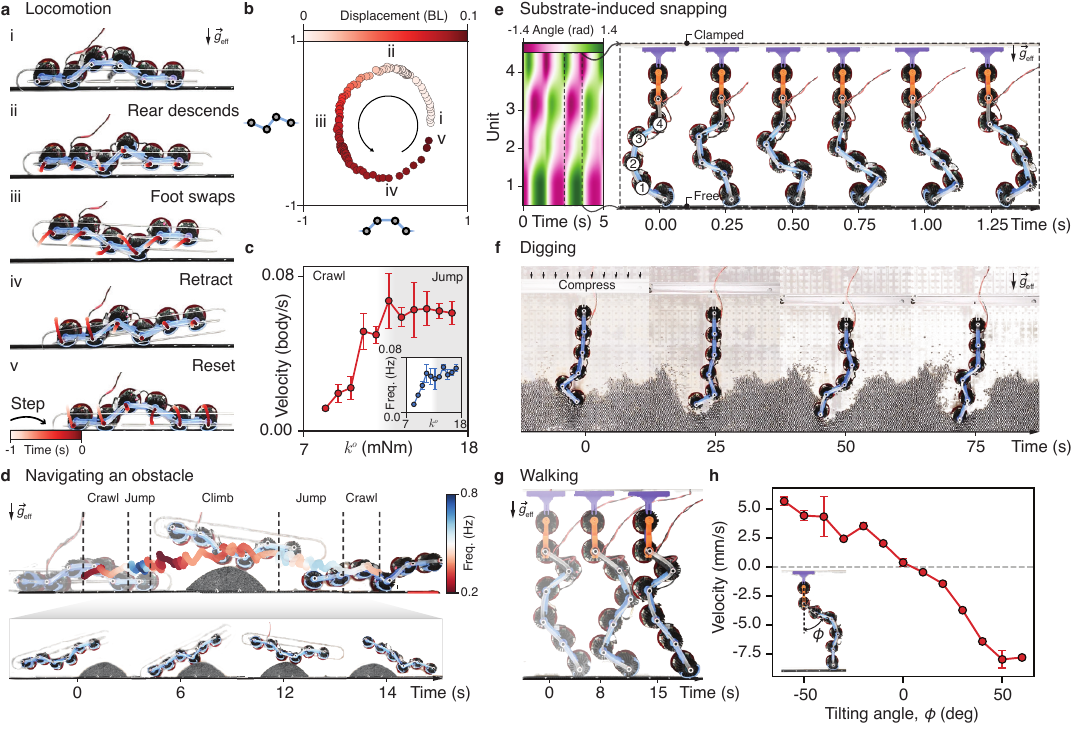}
    \caption{\textbf{Polyfunctions of non-reciprocal filaments.}  The same filament exhibits different locomotion modalities in response to distinct environmental perturbations.
    \textbf{(a, b)} Buckling our filaments via a free-standing brace, unidirectional shape cycles enable crawling on a substrate. Key frames are shown in panel (a) (i-v), with projections onto the lowest shape modes in (b). First, the rear linkage descends, driving the chain forward (i, ii). The filament's anchor point then swaps to the front linkage (iii), as the rear retracts (iv) and the cycle repeats (v). Displacement measured in body lengths (BL).
    \textbf{(c)} Locomotion velocity and snapping frequency versus the non-reciprocity $k^{o}$. The filament switches gait from crawling to jumping as the snapping frequency reaches the timescale of vertical motion in the filament's centre of mass.
    \textbf{(d)} The active filament navigates an obstacle by spontaneously switching gaits. The trajectory of the centre of mass is shown here and colored by the snapping frequency. As the filament touches the obstacle and leaves the substrate, it switches to the jumping gait. Once the middle units touch the obstacle, the filament snaps slowly again and lifts its body to climb up the obstacle. Inset shows details of the transition where the filament jumps, climbs, then jumps over the obstacle. 
    \textbf{(e)} Driving a half-clamped, half-unconstrained filament into a substrate, the substrate itself causes buckling and hence repeated snapping. However, now the bottom-most node of the chain can additionally translate or leave the substrate. Inset shows a kymograph with the angle deviations at free vertices over time, which clearly shows unidirectional waves.
    \textbf{(f)} A standing active filament digs into a granular pile of steel beads. While compressing the active filament, it scoops steel beads and pulls them aside as it snaps.
    \textbf{(g)} Breaking left–right symmetry enables the standing filament to walk. We tilt the filament by an angle $\phi$ and allow the top end to slide freely. The filament walks persistently towards the tilting direction. 
    \textbf{(h)} Walking velocity versus the tilt angle $\phi$.
    }  
    \label{fig:Locomotion}
\end{figure*}

We have seen that buckling filaments at sufficiently high $k^\mathrm{o}$ drives them into a limit cycle of shape changes. We now use this mechanistic insight to enable multiple modalities of locomotion in a single filament. We first buckle our chain with a brace that constrains its length while allowing it to translate freely in space.
We place the truss horizontally on a substrate under gravity [Fig.~\ref{fig:Locomotion}(a)]. During a single shape-change cycle, the two freely moving nodes of the truss now contact the ground. These ground contacts exert normal forces on the beam, lifting it off the substrate, alongside tangential traction forces which drive the filament forward in a crawling motion [Fig.~\ref{fig:Locomotion}(a), Supplementary Video 4]. The direction of motion is aligned with the sense of wave amplification: The beam crawls via a wave that originates from its rear and propagates to its head, analogously to the metachronal waves of the caterpillar~\cite{brackenburyCaterpillarKinematics1997} or the lagged phase of microswimmers~\cite{purcellLifeLowReynolds1977a, ishimotoSelforganizedSwimmingOdd2022}. Despite the environmental perturbation of the substrate, limit cycles of snapping remain an attractor for the beam dynamics [Fig.~\ref{fig:Locomotion}(b)]: In Methods Fig.~\ref{fig:Robustness} and Supplementary Video 5, we repeatedly manually disrupt our crawler, which returns to the same cycle after each perturbation.

Crawling speed increases with $k^\mathrm{o}$ until the periodic motion of the chain's center of mass becomes comparable to its intrinsic snapping frequency [Fig.~\ref{fig:Locomotion}(c)]. At this point we find that the filament changes gait~\cite{heydariSeaStarInspired2020}: both limbs leave the substrate simultaneously, and the filament switches from crawling to bouncing. This bouncing mode is automatically invoked whenever the filament climbs an obstacle with dimensions matching its own body size. As motors leave the substrate during climbing they oscillate rapidly, providing a source of noise that aids the filament in overcoming local minima during its ascent [Fig.~\ref{fig:Locomotion}(d), Supplementary Video 4].

Our crawler is powered by an external brace, which maintains the filament in a buckled state. We now allow the environment itself to cause buckling, driving a feedback loop between filament snapping and environmental perturbation that will autonomously select a locomotion gait. We clamp a filament at one end, and drive the unconstrained end into a substrate. The substrate now mimics the missing constraint by exerting an opposing normal force on the bottom-most node of chain, and the beam buckles and begins to snap [Fig.~\ref{fig:Locomotion}(e), Supplementary Video 4]. However, the substrate can only exert one-sided normal forces and finite tangential traction forces on the lowest node. This soft constraint allows the lowest node to slide relative to the substrate or leave it entirely, enabling a range of persistent scooping and stepping beam motions. Which motion the beam adopts will vary with the properties of the substrate, i.e. the traction and normal forces that the substrate exerts on the beam.

Driving our half-constrained chain into a granular pile of steel beads, we find that the chain snapping digs a hole in the pile [Fig.~\ref{fig:Locomotion}(f), Methods \S 4]. Excavation ends when the filament has dug enough space to return to its rest length. Because we have broken the symmetry of the boundary conditions applied to the chain, we also induce an additional polarity running from the clamped end to the free end. This polarity breaks the symmetry $k^\mathrm{o} \rightarrow -k^\mathrm{o}$: A wave amplifying from the clamped end to the free end results in a different gait to a wave amplifying from free to clamped. When waves run into the granular pile, we find a burrowing gait and a narrower hole. By contrast, when the wave begins within the pile, we find larger sweeping motions and a wider excavation [Supplementary Video 4].

The interplay between buckling via an external load and unidirectional beam snapping suggests that our active filaments would make natural walkers. Indeed, when we allow the vertical beam constraint shown in Fig.~\ref{fig:Locomotion}(e) to freely slide horizontally, the traction forces on the lowest node cause the centre-of-mass of the beam to repeatedly translate back-and-forth [Supplementary Video 4]. Just as for the crawling motion of our truss, this stepping motion is driven by an attractive cycle of shape change that is robust to manual perturbation [Methods Fig.~\ref{fig:Robustness}, Supplementary Video 5]. However, despite breaking $k^\mathrm{o} \rightarrow - k^\mathrm{o}$ symmetry along the beam, the geometry shown in Fig.~\ref{fig:Locomotion}(e) preserves a left/right reflection symmetry perpendicular to the beam, prohibiting net translation---the walker takes a step forward, but an equal step back.

However, upon tilting the filament relative to the substrate by an angle $\phi$ we break this symmetry and immediately find an asymmetric walking gait and net translation [Fig.~\ref{fig:Locomotion}(g), Supplementary Video 4], with a walking velocity which scales as $\phi$ [Fig.~\ref{fig:Locomotion}(h)]. Analogously to the digger, the gait of the walker depends on sense of wave propagation, and we find that only waves leading from the foot result in a persistent walking gait. 

We have demonstrated that buckling in materials which break reciprocity enables a distinctive class of instabilities: not only snapping between static states, but also self-sustained oscillations. These oscillations occur in free-standing structures, in contrast to previous work on active filaments \cite{sekimotoSymmetryBreakingInstabilities1995,decanioSpontaneousOscillationsElastic2017, filyBucklingInstabilitiesSpatiotemporal2020,clarkeBifurcationsNonlinearDynamics2024,zhengSelfOscillationSynchronizationTransitions2023}, a crucial feature that we exploit to engineer self-supported locomotors which crawl, dig, and walk.

The oscillations of our active beams are distinct from the linear vibrations of a non-Hermitian system driven across an exceptional transition, because they arise from limit cycles in the nonlinear dynamics of the beam. Our theory captures how this process unfolds. At the linear level, non-reciprocal couplings cause multiple eigenmodes of the beam to become degenerate, and buckling drives these coupled modes unstable---this is the critical exceptional point (CEP). The internal nonlinearity of beam buckling then tames the post-bifurcation dynamics of these linearly unstable coupled modes into a finite-amplitude stable limit cycle [Fig.~\ref{fig:Truss}(g)]. This limit-cycle dynamics persists when our beams encounter environmental perturbations, enabling our locomotors to continue functioning after repeated disturbances [Figs.~\ref{fig:Truss}(c,d), Fig.~\ref{fig:Locomotion}(d), Fig.~\ref{fig:Robustness}].

We have focused on the lowest modes of our active beams [Fig.~\ref{fig:buckling}(b)], and captured the key features of their dynamics using a von-Mises truss built from only four nodes (Fig.~\ref{fig:Truss}). Longer chains exhibit this same snapping dynamics for large viscous dissipation $\Gamma$. The highly viscous limit that we operate in here can also be considered the poor oscillator limit $Q\rightarrow 0$, where the quality factor $Q$ of a torsional oscillator is given by $Q^2 = {m B l_0^2}/{\Gamma^2}$. Recent observations of multimode interactions in unconfined active chains~\cite{veenstraWaveCoarseningDrives2025} suggest that at larger $Q$---achieved through reduced dissipation $\Gamma$ or increased mass $m$---longer buckled beams may support additional modes that can be selectively excited for further states of locomotion. To model the locomotion of such extended chains a natural method is discrete differential geometry (DDG), extended to a non-variational dynamics that builds non-reciprocal couplings directly into the stiffness matrix~\cite{tongDiscreteDifferentialGeometry2026, huangTutorialSimulatingNonlinear2025}. DDG would prove particularly useful for interactions with a complex substrate [Figs.~\ref{fig:Locomotion}(d,f)] or surrounding fluid.

The snapping dynamics of a passive beam is well-studied~\cite{pandeyDynamicsSnappingBeams2014}. Elastic snap-through occurs through a family of bifurcations---pitchforks, saddle-nodes---which depend on the symmetries of the problem~\cite{gomezCriticalSlowingPurely2017,  liuDelayedBifurcationElastic2021, radissonElasticSnapthroughInstabilities2023}. For example, symmetry-breaking tilts imposed at either end of the beam can act as a bifurcation parameter. By contrast, we have focused on simple symmetric loading, and our bifurcation parameter has been internal non-reciprocal coupling. Combining these two parameters might enable fine control over snapping dynamics, or tuneable switching between oscillations and stable states. Indeed, passive beam buckling has underpinned a generation of structures and architected materials that achieve functionality through nonlinearity~\cite{melanconInflatableOrigamiMultimodal2022, nadkarniUnidirectionalTransitionWaves2016, leeMechanicalNeuralNetworks2022, kwakernaakCountingSequentialInformation2023}.
We anticipate that active buckling represents a similarly fruitful pathway towards unconventional soft robots and multistable active matter.

\section*{Materials and Methods}
\section{The continuum mechanics of a non-reciprocal beam}
\label{sec:Continuum}
Here we describe the derivation of the odd stress~\eqref{eq:EoM}, the close-to-flat non-reciprocal beam equation~\eqref{eq:NRBeam}, and the stability analysis~\eqref{eq:RingDispersion}. In the Supplementary Materials we also provide a description of the beam kinematics for arbitrary geometries.

The simplest odd stress we can write down that breaks left-right symmetry must depend on even derivatives of the curvature (as if the stress depends on odd derivatives then it will give a force that is even in symmetry). This stress is of the form

\begin{equation}
    {\boldsymbol \sigma}^\text{odd}= \zeta \partial_s^{2}\kappa {\bf t}{\bf n}\text{,}
\end{equation}

where $\bf t$ is the tangent to the beam, and $\bf n$ the normal. Upon taking the divergence with $\nabla = {\bf t}\partial_s$ we find 
\begin{equation}
    {\bf f} = \nabla\cdot{\boldsymbol \sigma^{\mathrm{odd}}}  = \zeta \partial_s^{3} \kappa {\bf n} + O(\kappa^2). 
    \label{eq:Fodd2}
\end{equation}
The force~\eqref{eq:Fodd2} appears as~\eqref{eq:EoM} in the Main Text, and advects even derivatives of the curvature along the filament material frame. For a close-to-flat filament, ${\bf x} = (x(s,t),h(s,t))$ where $h(s,t)\ll 1$. At zeroth order in height deformations the 
curvature is given by $\kappa = - h_{xx}$, and arclength derivatives $\partial_s$ become derivatives along $x$, $\partial_x$. In this limit, the non-reciprocal forces~\eqref{eq:Fodd2} reduce to $\zeta \partial^5_x h$. We add this term to the dynamic beam equation under viscous damping~\cite{kodio_thesis, Buckmaster_Nachman_Ting_1975} to give
\begin{equation}
     \beta \dot{h} =  \Lambda h_{xx} - A h_{xxxx} - \zeta h_{xxxxx} - \eta\dot{h}_{xxxx},
    \label{eq:LinearizedTheory}
\end{equation}

where $\beta$ is substrate friction, $\Lambda$ is line tension, $A$ is bending rigidity and $\eta$ is bending viscosity.
Taking the ansatz $h(x,t)=e^{(\lambda t + \mathrm{i} q x)}$ we obtain the dispersion
\begin{align}
    \lambda(q) = \frac{-1}{\beta+\eta q^4} \left(\Lambda q^2 + A q^4 + \mathrm{i}\zeta q^5  \right)\text{,}
    \label{eq:DispersionSI}
\end{align}

where $\lambda$ is the growth rate of the spatial Fourier mode with wavenumber $q$.Equation~\eqref{eq:DispersionSI} appears as~\eqref{eq:RingDispersion} of the Main Text.

\section{Analysis of the odd von Mises truss } \label{sec:TrussTheory}
To give a deeper understanding of the snapping behavior of the odd beam we consider the example of a von Mises truss~\cite{howellCompliantMechanisms, wangTransientAmplificationBroken2024}, with the addition of an odd active torque. 

\subsection{Derivation of the equations of motion via the principle of virtual work}

The linkage is defined by the four vertices, $\{\text{L},\text{LC},\text{RC},\text{R}\}$, and associated internal angles, left $\theta_\text{L}$, left-central $\theta_\text{LC}$, right-central $\theta_\text{RC}$ and right $\theta_\text{R}$, along with the positions of the two central points connected by springs with rest length $l_0$. The beam is flat when $\theta_\text{L}=\theta_\text{R}=0$ and $\theta_\text{LC}=\theta_\text{RC}=\pi$. See Fig.~3 in the Main Text for a schematic.

The odd active torque for each of the vertices is given by
\begin{align}
    \tau_\text{L} &= k^\text{o} (\theta_\text{LC}-\pi)\text{,}\\
    \tau_\text{LC} &= k^\text{o} \left(\theta_\text{RC}-\pi- \theta_\text{L} \right)\text{,}\\
    \tau_\text{RC} &= k^\text{o} \left(\theta_\text{R}- \theta_\text{LC} + \pi \right)\text{,}\\
    \tau_\text{R} &= -k^\text{o} (\theta_\text{RC}-\pi)\text{,}
\end{align}
where $k^\text{o}$ is the microscopic non-reciprocity. In addition to these active torques the bar has a bending energy given by
\begin{align}
    E_b = \frac{B}{2}\left[ \theta_\text{L}^2 + \theta_\text{R}^2 + \left( \theta_\text{LC}-\pi\right)^2 + \left( \theta_\text{RC}-\pi\right)^2 \right]\text{,}
    \label{eq:Bend}
\end{align}
with $B$ the microscopic bending stiffness. The stretch energy for the system is given by
\begin{align}
    E_s =&\frac{1}{2} k \bigg(l_\text{0}-\big[(-\text{d}l+l-w_\text{L}-w_\text{R})^2\nonumber\\
    &\qquad+(w_\text{L} \tan (\theta_\text{L})- w_\text{R} \tan (\theta_\text{R}))^2\big]^{1/2}\bigg)^2\nonumber\\
    &+\frac{1}{2} k \left(l_0-w_\text{L} \sec (\theta_\text{L}))^2+(l_0-w_\text{R} \sec (\theta_\text{R}))^2\right)\text{,}
    \label{eq:Stretch}
\end{align}
where $w_\text{L,R}$ are the distances of the inner vertices projected onto the line connecting the fixed outer vertices, $\text{d}l$ is the compression distance and $l$ the total length of the uncompressed truss.

We will consider the limit where viscous dissipation within the hinges dominates the truss dynamics. We derive our equations of motion in this regime using a Rayleigh dissipation function $\mathcal{R}(\theta, \dot{\theta})$, which includes all terms contributing to a rate of work generation or dissipation within the truss~\cite{doiSoftMatterPhysics2013, Doi2011}:
\begin{equation}
\begin{aligned}
    &\mathcal{R} = \frac{\Gamma}{2}\left[ \dot\theta_\text{L}^2 + \dot\theta_\text{LC}^2 + \dot\theta_\text{RC}^2 + \dot\theta_\text{R}^2\right] +\\
    &k^\text{o}[ (\theta_\text{LC} -\pi)\dot\theta_\text{L}
    +  (\theta_\text{RC}-\pi-\theta_\text{L}) \dot\theta_\text{LC}
    \\+ &(\theta_\text{R}-\theta_\text{LC}+\pi) \dot\theta_\text{RC}- (\theta_\text{RC}-\pi) \dot\theta_\text{R}]\text{.}
    \label{eq:Rayleigh}
\end{aligned}
\end{equation}
Here $\Gamma$ is the hinge dissipation and $k^\text{o}$ the microscopic non-reciprocity. Terms like $\frac{\Gamma}{2} \dot{\theta_{L}}^2$ in~\eqref{eq:Rayleigh} are the viscous power dissipated by hinge $\theta_L$. Terms like $k^\mathrm{o} (\theta_\text{LC} -\pi)\dot\theta_\text{L}$ are the power generated by the odd torques acting on vertex $\theta_\text{L}$. Given a Lagrangian $L(\theta, \dot{\theta})$ that encodes conservative dynamics, and a dissipation function $\mathcal{R}(\theta, \dot{\theta})$, the Rayleighan formalism gives equation of motion via
\begin{align}
\frac{d}{dt} \left( \frac{\partial L }{\partial \dot{\theta}} \right)-
\frac{\partial L }{\partial \theta}=
-\frac{\partial \mathcal{R} }{\partial \dot{\theta}}.
\label{eq:Rayleighan}
\end{align}
In the truss, the Lagrangian $L=-(E_s+ E_b)$ where the potential energy is the sum of bending and stretching energies~\eqref{eq:Bend},~\eqref{eq:Stretch}, neglecting inertia. The dissipation $\mathcal{R}$ is given by~\eqref{eq:Rayleigh}. 

Before deriving the equations of motion, we first make a series of small angle approximations to simplify $E_b$, $E_s$ and $\mathcal{R}$. We will expand about a nearly flat state, introducing a small parameter $\epsilon$ that captures this small-strain approximation. More precisely, we will see that the compression $\text{d}l \sim O(\epsilon^2)$. 
We first rewrite $\theta_{\text{RC},\text{LC}}$ and their time derivatives using the geometric relations
\begin{widetext}
\begin{align}
    \theta_\text{LC} &= \pi - \theta_\text{L}
    -\text{arctan}\left(-\text{d}l+l-w_\text{L}-w_\text{R},w_\text{L} \tan (\theta_\text{L})-w_\text{R} \tan (\theta_\text{R})\right)\text{,}\nonumber\\
    \theta_\text{RC} =& \pi - \theta_\text{R}
    +\text{arctan}\left(-\text{d}l+l-w_\text{L}-w_\text{R},w_\text{L} \tan (\theta_\text{L})-w_\text{R} \tan (\theta_\text{R})\right)\nonumber\text{.}
\end{align}
\end{widetext}
We can then define the following symmetrized/antisymmetrized variables and expand in a small parameter $\epsilon$ (\textit{i.e.}~expand in small deflections from a flat state):
\begin{align}
    &\theta_\text{L} = \epsilon\left(\theta_\text{S} + \frac{1}{2} \theta_\text{A}\right)\text{,}\\
    &\theta_\text{R} = \epsilon\left(\theta_\text{S} - \frac{1}{2} \theta_\text{A}\right)\text{,}\\
    & w_\text{L}= l_{0}-\frac{1}{2} \epsilon ^2 (\mathrm{d}l+w_\text{A}-w_\text{S})\text{,}\\
    & w_\text{R}= l_{0}-\frac{1}{2} \epsilon ^2 (\mathrm{d}l-w_\text{A}-w_\text{S})\text{,}
\end{align}
where we also assume that $\text{d}l$ is $O(\epsilon^2)$, and rescale as $\text{d}l\sim\epsilon^2 \text{d}l$. 

We also non-dimensionalize energies by $B$ and lengths by $l=3l_0$.
Time will be renormalized by $\Gamma/B$. Thus our dimensionless parameters are given by
\begin{align}
    &k^\text{o}\leftarrow \frac{k^\text{o}}{B}\text{,}\\
    & k \leftarrow \frac{k l^2}{B}\text{,}\\
    & \text{d}l \leftarrow \frac{\text{d}l}{l}\text{.}
\end{align}

Expanding to $4^\text{th}$ order in $\epsilon$ gives the following energy:

\begin{align}
    &E_b+E_s =\frac{1}{2} \epsilon ^2 \left(5 \theta_A^2+4 \theta_S^2\right)\nonumber\\
    &+ \frac{1}{576} \epsilon ^4 \Big(144 \text{d}l^2 k-2592 \text{d}l \theta_A^2-24 \text{d}l \theta_A^2 k-96 \text{d}l \theta_S^2 k\nonumber\\
    & \qquad -288 \text{d}l k w_\text{S}-432 \theta_A^4+1728 \theta_A^2 \theta_S^2+9 \theta_A^4 k\nonumber\\
    & \qquad +24 \theta_A^2 \theta_S^2 k+16 \theta_S^4 k+144 k w_\text{A}^2-96 \theta_A \theta_S k w_\text{A}\nonumber\\
    &\qquad+432 k w_\text{S}^2-72 \theta_A^2 k w_\text{S}+96 \theta_S^2 k w_\text{S}-5184 \theta_A \theta_S w_\text{A}\nonumber\\
    &\qquad+7776 \theta_A^2 w_\text{S}\Big)\text{,}
\end{align}
and the dissipation is given by
\begin{align}
  &\mathcal{R} = \frac{1}{2} \epsilon ^2 \left(10 k^\text{o} \theta_\text{S} \dot\theta_\text{A}-10 k^\text{o} \theta_\text{A} \dot\theta_\text{S}+5 \dot\theta_\text{A}^2+4 \dot\theta_\text{S}^2\right)\nonumber\\
  &\qquad+  \frac{1}{4} \epsilon ^4 \Big(-72 \text{d}l k^\text{o} \theta_\text{S} \dot\theta_\text{A}+72 \text{d}l k^\text{o} \theta_\text{A} \dot\theta_\text{S}-54 \text{d}l \dot\theta_\text{A}^2\nonumber\\
  &\qquad-12 k^\text{o} \theta_\text{A}^2 \theta_\text{S} \dot\theta_\text{A}+16 k^\text{o} \theta_\text{S}^3 \dot\theta_\text{A}+4 k^\text{o} \theta_\text{A}^3 \dot\theta_\text{S}+16 k^\text{o} \theta_\text{A} \theta_\text{S}^2 \dot\theta_\text{S}\nonumber\\
  &\qquad-48 k^\text{o} \theta_\text{S}^2 \dot w_\text{A}+72 k^\text{o} \theta_\text{A} \theta_\text{S} \dot w_\text{S}+72 k^\text{o} \theta_\text{S} w_\text{S} \dot\theta_\text{A}\nonumber\\
  &\qquad-72 k^\text{o} \theta_\text{A} w_\text{S} \dot\theta_\text{S}+24 \theta_\text{A} \theta_\text{S} \dot\theta_\text{A} \dot\theta_\text{S}+12 \theta_\text{S}^2 \dot\theta_\text{A}^2-9 \theta_\text{A}^2 \dot\theta_\text{A}^2\nonumber\\
  &\qquad-36 \theta_\text{S} \dot\theta_\text{A} \dot w_\text{A}-36 w_\text{A} \dot\theta_\text{A} \dot\theta_\text{S}+54 \theta_\text{A} \dot\theta_\text{A} \dot w_\text{S}+54 w_\text{S} \dot\theta_\text{A}^2\Big)\text{.}
  \label{eq:RayleighFourthOrder}
\end{align}
We assume that the spring lengths relax much faster than the angular degrees of freedom, and make the quasistatic approximation that these springs are always at equilibrium. This approximation amounts to slaving $w_\text{S,A}$ to $\theta_\text{S,A}$ by minimizing the energy with respect to $w_\text{S,A}$: 
\begin{align}
\frac{\partial}{\partial w_\text{{S,A}}}(E_b+E_s) =0,
\end{align}
which gives
\begin{align}
    &w_\text{A} = \frac{(k+54) \theta_\text{A} \theta_\text{S}}{3 k},\\
    &w_\text{S} = \frac{1}{9} \left(3 \text{d}l-\theta_\text{S}^2\right)+\left(\frac{1}{12}-\frac{9}{k}\right) \theta_\text{A}^2\text{.}
\end{align}
We then set $\dot w_\text{S,A}=0$ in the dissipation functional~\eqref{eq:RayleighFourthOrder} and calculate the dynamical equations for only the two angular degrees of freedom. These equations are found by application of the variational principle~\eqref{eq:Rayleighan} as
\begin{equation}
    \frac{\partial E}{\partial \theta_\text{S,A}} + \frac{\partial \mathcal{R}}{\partial \dot\theta_\text{S,A}} = 0\text{,}
\end{equation}
which to third order in $\epsilon$ gives
\begin{widetext}
\begin{align}
    \dot\theta_{S} &= 
     \left(\frac{dl k}{18}-1\right)\theta_S
    +\left(\frac{5}{4}-3 {dl}\right)k^\mathrm{o}\theta_A 
    + \left(\frac{3}{2}-\frac{k}{72}\right)\theta_A^2 \theta_S\nonumber\\
    & \qquad+ \left(-\frac{81 {}}{2 k}-\frac{3 {}}{4}\right)k^\mathrm{o}\theta_A \theta_S^2 
    + \left(\frac{{1}}{8}-\frac{81 {}}{2 k}\right)k^\mathrm{o}\theta_A^3 
    -\frac{k}{54}\theta_S^3 \nonumber\\
    \dot\theta_A &= 
 \left(\frac{{dl} k}{30}-\frac{18 {dl}}{5}-1\right)\theta_A
+ \left(-\frac{6 {dl} {}}{5}-{1}\right)k^\mathrm{o}\theta_S
+\left(\frac{9}{5}-\frac{k}{90}\right)\theta_A\theta_S^2 \nonumber\\
& \qquad+ \left(\frac{243 }{10 k}-\frac{9 }{10}\right)k^\mathrm{o}\theta_A^2 \theta_S
+ \left(-\frac{k}{120}-\frac{3}{4}\right)\theta_A^3
+\frac{{k^\mathrm{o}}}{5}\theta_S^3.
\label{eq:EoMSI}
\end{align}
\end{widetext}
Equation~\eqref{eq:EoMSI} is of the general cubic form given in~\eqref{eq:TrussDynamics} of the Main Text, namely
\begin{equation}
    \dot{\boldsymbol{\theta}} = J\boldsymbol{\theta} + \sum_{j=0}^{3}
         {\bf A}_{m} \theta_\text{S}^m \theta_\text{A}^{3-m} \text{,}
\label{eq:EoMSI2}
\end{equation}
where $\boldsymbol{\theta}=(\theta_\mathrm{S},\theta_\mathrm{A})$. The coefficients $J$ and $\bf A$ read 
\begin{equation}
\begin{aligned}
    &J = \begin{pmatrix}
        \frac{dl k}{18}-1 & \left(\frac{5}{4}-3 {dl}\right)k^\mathrm{o}\\
        \left(-\frac{6 {dl} {}}{5}-{1}\right)k^\mathrm{o} & \frac{{dl} k}{30}-\frac{18 {dl}}{5}-1
    \end{pmatrix}\text{,}\\
    & {\bf A}_0 = \begin{pmatrix}
         \left(\frac{{1}}{8}-\frac{81 {}}{2 k}\right)k^\mathrm{o} \\
        -\frac{k}{120}-\frac{3}{4}
    \end{pmatrix}\text{,}\\
    & {\bf A}_1 = \begin{pmatrix}
        \frac{3}{2}-\frac{k}{72}\\
       \left( \frac{243 }{10 k}-\frac{9 }{10} \right) k^\mathrm{o}
    \end{pmatrix}\text{,}\\
    & {\bf A}_2 = \begin{pmatrix}
        \left(-\frac{81 {}}{2 k}-\frac{3 {}}{4}\right)k^\mathrm{o}\\
        \frac{9}{5}-\frac{k}{90}
    \end{pmatrix}\text{,}\\
    & {\bf A}_3 = \begin{pmatrix}
        -\frac{k}{54}\\
        \frac{k^\mathrm{o}}{5}
    \end{pmatrix}\text{.}
    \label{eq:Coefficients}
\end{aligned}
\end{equation}
As described in the Main Text, we note that the $\mathbb{Z}_2$ symmetry $(\theta_S, \theta_A) \rightarrow (-\theta_S, -\theta_A)$ forces nonlinear terms of cubic order only in~\eqref{eq:EoMSI2}. The symmetry under simultaneous reflection and a reversal in the sense of non-reciprocity, $(\theta_S, \theta_A, k^\mathrm{o} )\rightarrow (\theta_S, -\theta_A, -k^\mathrm{o} )$ constrains $k^\mathrm{o}$ to appear only in the off-diagonal elements of $J$ in~\eqref{eq:Coefficients}, and enforces the alternating pattern of terms proportional to $k^\mathrm{o}$ in ${\bf A}_i$. More formally, we split these terms into ${\bf A}_i = {\bf A}_i^\mathrm{passive}(k) + {\bf A}_i^\mathrm{odd}(k^\mathrm{o},k)$. We require the symmetry ${\bf A}^\mathrm{odd}_i=(-1)^{i} \sigma_z{\bf A}^\mathrm{odd}_i$, with $\sigma_z$ a Pauli matrix. We now analyze the bifurcation structure of the dynamics ~\eqref{eq:EoMSI2}.

\subsection{Bifurcation structure of the odd von Mises truss}
Figures~\ref{fig:Truss}(f, g) show a selection of phase portraits for the truss dynamics~\eqref{eq:EoMSI2} as the non-reciprocity $k^\mathrm{o}$ and compression $\text{d} l$ are varied. In the limit of no active force, $k^\text{o}=0$, we find classic symmetrically buckled solutions
\begin{equation}
    \theta_\text{S}^\star=\pm \frac{\sqrt{3} \sqrt{\text{d}l k-18}}{\sqrt{k}}, \quad \theta_\text{A}^\star=0\text{.}
\end{equation}
As $k^\mathrm{o}$ is increased, these buckled fixed points merge in a SNIC bifurcation, and a limit cycle is born. 

To locate the critical non-reciprocity $k_c^o$ and compression $\text{d}l_c$ for the CEP shown in Fig.~3 of the Main Text, we demand a double zero in the eigenvalues of the Jacobian $J$ in~\eqref{eq:Coefficients}. We enforce this condition by solving for $\text{Tr}({J})=\text{Det}({J})=0$, which gives
\begin{align}
    &k^\text{o}_\text{c} = \pm \frac{k+162}{\sqrt{20 k^2-2160 k+25515}},\\
    &\text{d}l_\text{c} = \frac{45}{2 k-81}\text{.}
\end{align}
In the incompressible limit $k\rightarrow \infty$ these expressions simplify to 
\begin{align}
    &k^\text{o}_\text{c} = \pm \left(\frac{108}{\sqrt{5}k} + \frac{1}{2\sqrt{5}} \right),\\
    &\text{d}l_\text{c} = \frac{45}{2 k}\text{,}
    \label{eq:ThresholdSI}
\end{align}
which is the location of the CEP given in the Main Text. Substituting these critical values back into the Jacobian~\eqref{eq:Coefficients}, perturbing as $k^\text{o}= k^\text{o}_\text{c} + \delta k^\text{o}$, $\text{d}l = \text{d}l_\text{c} + \delta l/k$ and taking the limit $k\to\infty$ we find the normal form
\begin{equation}
    {J}_{\text{CEP}} = \left(
\begin{array}{cc}
 \frac{1}{36} (2 \delta l  +9) & \frac{1}{8} \left(10 \delta k^\text{o}+\sqrt{5}\right) \\
 -\delta k^\text{o}-\frac{1}{2 \sqrt{5}} & \frac{\delta l }{30}-\frac{1}{4} \\
\end{array}
\right)\text{,}
\end{equation}
which is given as~\eqref{eq:CEP} of the Main Text. In order to find the theoretical snapping frequencies shown in Fig.~\ref{fig:Truss}(e) of the Main Text, we look for the imaginary part of eigenvalues, $\lambda_{J}$ of $J$ in the vicinity of the CEP:
\begin{equation}
    \lambda_{J} = \frac{1}{90} \left(4 \delta l \pm\sqrt{\delta l  (\delta l +45)-2025 \delta k^\text{o} \left(5 \delta k^\text{o}+\sqrt{5}\right)}\right)
    \label{eq:EigenValues},
\end{equation}
which gives the scaling $\omega\sim \sqrt{ k^\text{o}-k^\text{o}_\text{c}}$ characteristic of an exceptional point.

\section{Construction of our robotic metamaterials} \label{sec:Construction}
The 1D robotic filaments shown in Figs.~\ref{fig:buckling}--\ref{fig:Locomotion} are composed of motorised vertices connected by plastic arms. Each vertex consists of a DC coreless motor (Motraxx CL1628) embedded in a cylindrical heatsink, an angular encoder (CUI AMT113S), and a microcontroller (ESP32) connected to a custom electronic board. The electronic board enables power conversion, interfacing between the sensor and motor, and communication between vertices.
Each vertex has a diameter $50$mm, height $90$mm, and mass $0.2$kg. 
The power necessary to drive the motor is provided 
by an external 48V DC power source.

\begin{figure}[t]
    \centering
    \includegraphics[width=\linewidth]{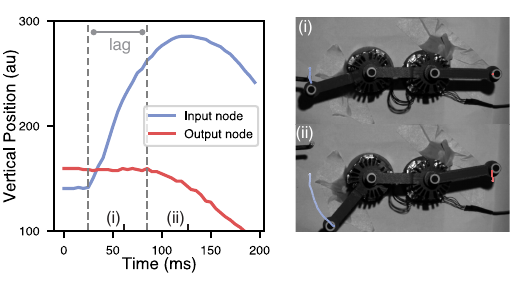}
    \caption{{\bf Response Time Calibration.} We estimate the timescale for the process of digital data acquisition, signaling, and motor actuation by applying an impulse to a 2-bar linkage, filmed under a high speed camera. We find a conservative upper bound of $\SI{50}{\milli \second}$, which is $40\times$ faster than the most rapid oscillations our beams.}
    \label{fig:ResponseTime}
\end{figure}

Rigid 3D-Printed arms connect each motor's drive shaft to the heat sink of the adjacent unit. The angle formed between the two arms at vertex $i$ is denoted by $\theta_i$. The on-board sensor measures $\theta_i$ at a sample rate of $100$Hz and communicates the measurement to nearest neighbours. In response to the incoming signal, vertex $i$ exerts an active torsional force $\tau_i^a$,
\begin{equation}
    \tau^a_{i} = \begin{cases}
        k^\mathrm{o} \left(\delta \theta_{i+1}-\delta \theta_{i-1} \right)\mathrm{, }|\delta \theta_{i+1}-\delta \theta_{i-1} |<\tau_{\mathrm{max}}/k^\mathrm{o}, \\
        \mathrm{sgn}(\delta \theta_{i+1}-\delta \theta_{i-1} )\tau_{\mathrm{max}}\mathrm{, }|\delta \theta_{i+1}-\delta \theta_{i-1} |>\tau_{\mathrm{max}}/k^\mathrm{o},
    \end{cases}
    \label{eq:TorqueAngle}
\end{equation}
where $\delta \theta_i = \theta_i - \theta^0$, $\theta^0$ is the rest angle and $\tau_\mathrm{max}$ is the torque at which the motors saturate. 
The microcontrollers attached to each motor process incoming angle data, communicate with their neighbors, and then signal motors to exert torques. As an estimate of this sense $\rightarrow$ communication $\rightarrow$ actuation process we take a 2-unit system, apply an impulse deflection to one node, and measure the output deflection at the other. This process is filmed under a high-speed camera at 200fps (Basler acA2040-90um). A conservative upper bound for this process is less than $\SI{50}{\milli \second}$ (Fig.~\ref{fig:ResponseTime}). This timescale is roughly $40\times$ faster than the most rapid oscillations our beams exhibit which have period $O(\SI{2}{s})$.

The torques and angles in~\eqref{eq:TorqueAngle} are implemented using arbitrary digital units---bits---that are specific to our robotic implementation. We calibrate these digital values against standard physical units by measuring torque-displacement slopes for different values of the electronic feedback, as described in Ref.~\cite{veenstraAdaptiveLocomotionActive2025}. We find a conversion factor $\SI{846}{\milli\newton\metre\per\radian\per\bit}$ taking us from digital to physical non-reciprocities $k^\mathrm{o}$. The coreless motor saturates at a maximum torque of $\tau_\mathrm{max}=\SI{12}{\milli\newton\metre}$. Each link has a length of $l_0=7.5\,\mathrm{cm}$. Adjacent vertices are also connected by viscoelastic hinges, whose materials properties we vary depending on the experimental protocol (\S\ref{sec:Protocol}). For further details, see Ref.~\cite{veenstraAdaptiveLocomotionActive2025}. 

\section{Experimental protocol}\label{sec:Protocol}
\begin{figure}[t]
    \centering
    \includegraphics[width=1\linewidth]{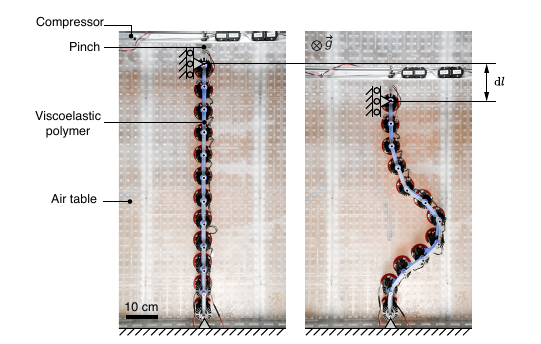}
    \caption{{\bf Details of our experimental buckling setup.} A typical setup during buckling experiments on our robotic metamaterials, showing a chain of 12 units clamped to an air table at either end, with an Agilus 30 viscoelastic polymer scaffold attached to the top of the chain to provide hinge elasticity and dissipation. Scale bar \SI{10}{\centi\metre}.}
    \label{fig:Setup_buckle}
\end{figure}
All experiments take place on top of a custom-made low-friction air table to minimize stick-slip substrate friction. The table consists of two $1.5\times 1.5$m Plexiglass plates that sandwich 5-mm-wide air channels. The top plate is pierced by an array of holes (pitch 10 mm, diameter 1 mm) conveying air pressurized at 10 bar. Each motorized vertex floats on a thin layer of pressurized air without making contact with the table. The experiments are filmed using a Nikon D5600 camera equipped with a 50-mm lens, recording 2-Mpx images at 50 frames per second. A marker is placed on the top of each vertex and we use the Python package OpenCV to detect and track the position of each vertex with a spatial resolution of 0.8 mm. We show a typical table setup in Fig.~\ref{fig:Setup_buckle}. 

To provide torsional rigidity and viscous dissipation to the hinges of our robotic metamaterials, we use two different viscoelastic polymers. For the data shown in Fig.~3, we use a silicone rubber with a thickness of $\SI{2}{\milli\metre}$. This rubber hinge is well described with a Kelvin-Voigt model, with passive elastic torsional stiffness of $B= \SI{20}{\milli\newton\metre\per\radian}$ and viscous dissipation $\Gamma =\SI{0.4\pm 0.1}{\milli\newton\metre\second\per\radian}$. See Ref.~\cite{veenstraAdaptiveLocomotionActive2025} for calibration details. The simple viscoelastic response of this rubber means that it is well suited for the detailed calibration performed in Fig.~\ref{fig:Truss}. However, for longer chains  (Figs.~\ref{fig:buckling},~\ref{fig:Locomotion}) we find that the dissipation of this rubber is insufficient to drive us into the viscous regime in which snapping is dominated only by the lowest modes of a longer chain.

\begin{figure}[t]
    \centering
    \includegraphics[width=\linewidth]{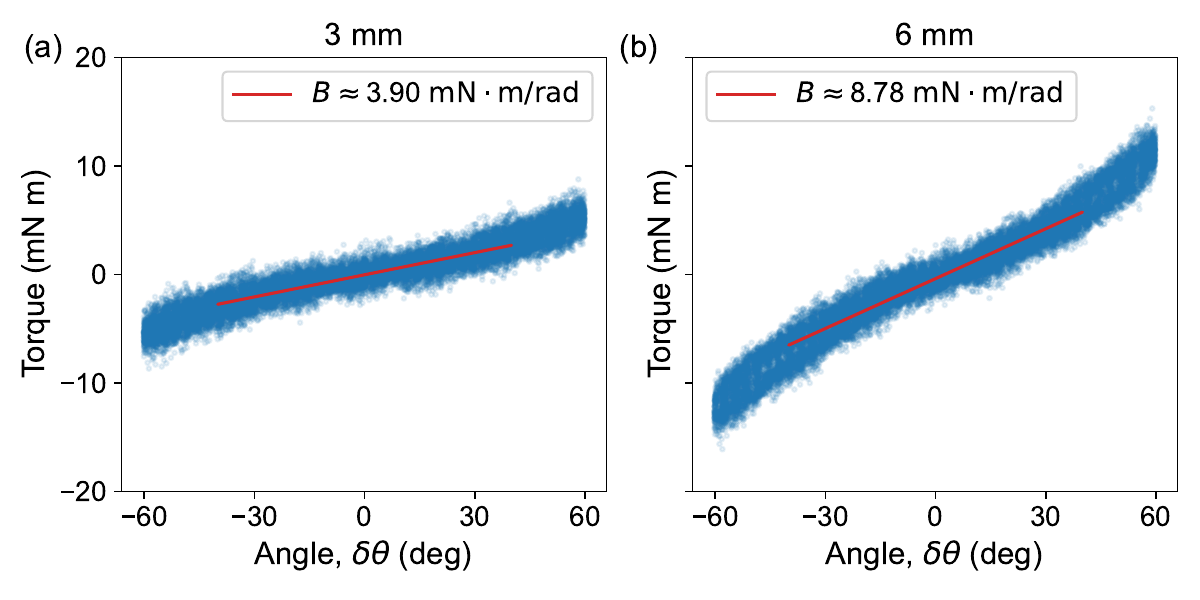}
    \caption{\textbf{Calibration of angular stiffness $B$ for our Agilus-30 hinges.} The restoring torque of the building block with a viscous Agilus band measured over a range of twist angles $\delta\theta$. The slope of the linear regression yields a value $B\approx 3.90\ \mathrm{mN\cdot m/rad}$ and $B\approx\ 8.78\ \mathrm{mN\cdot m/rad}$ for the rubber band with a thickness of (a) 3 mm and (b) 6 mm.}
    \label{fig:Agilus_kappa}
\end{figure}

\begin{figure}[t]
    \centering
    \includegraphics[width=\linewidth]{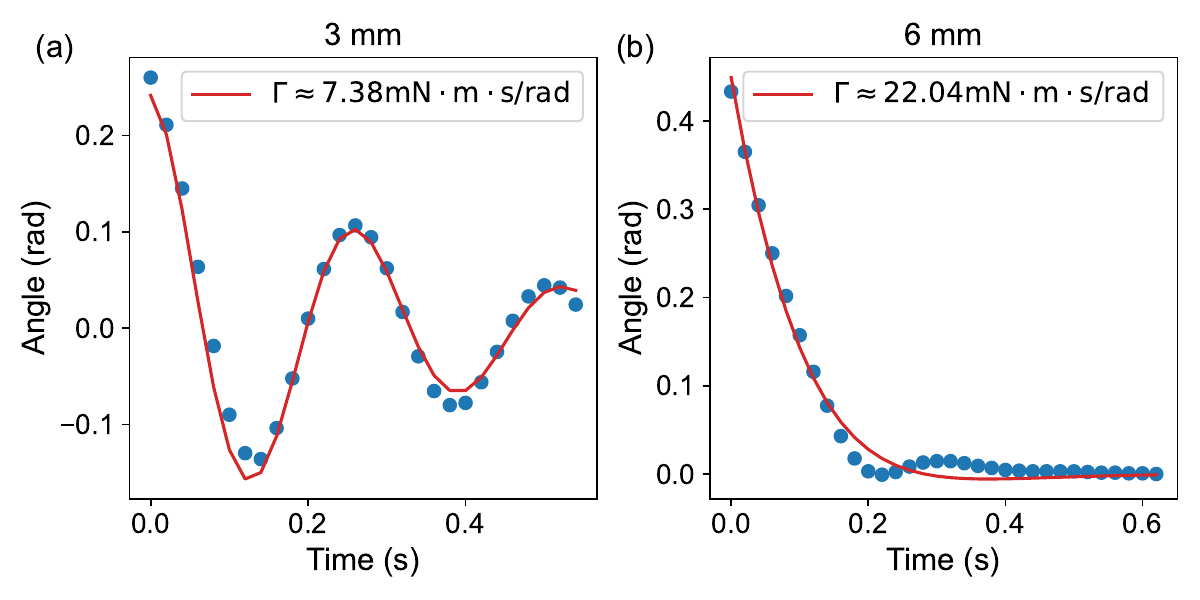}
    \caption{\textbf{Calibration of angular damping coefficient $\Gamma$ for our Agilus-30 hinges}. The decaying oscillations of the building block with a viscous Agilus band were fitted to the solution of a damped harmonic oscillator. We find viscous damping coefficients of $\Gamma\approx 7.38\ \mathrm{mN\cdot m \cdot s/rad}$ and $\Gamma\approx 22.04\ \mathrm{mN\cdot m \cdot s/rad}$ for the Agilus rubber band with a thickness of (a) 3 mm and (b) 6 mm.}
    \label{fig:Agilus_Gamma}
\end{figure}

For our chain snapping (Fig.~\ref{fig:buckling}) and locomotion (Fig.~\ref{fig:Locomotion}) experiments, we therefore use a highly viscous polymer. We 3D print hinges from a viscoelastic rubber-like PolyJet photopolymer (Stratasys Agilus 30). The viscoelastic response of this material has been characterised via a generalized Maxwell model in, for example, Ref.~\cite{dykstraViscoelasticSnappingMetamaterials2019}. Complex viscoelastic response makes the Agilus 30 ill-suited for the detailed calibration shown in Fig.~\ref{fig:Truss}. However, its increased viscous dissipation renders it ideal for our locomotion experiments. In this study, we used two thicknesses of Agilus 30 bands: 3 mm and 6 mm. We determined an approximate angular stiffness $B$ and angular dissipation $\Gamma$ by placing a single robotic vertex in a rotational testing device (Instron E3000 linear-torsional) equipped with a torsional load cell (angular resolution: 2.5 arcsecs; torque resolution: 1 mN m). To determine the angular stiffness $B$ of the rubber band, the torque was measured as a function of angular deviation. The value of $B$ is taken to be the slope of the linear fit. We find $B\approx 3.90\ \mathrm{mN\cdot m/rad}$ and $B\approx\ 8.78\ \mathrm{mN\cdot m/rad}$ for the rubber band with a thickness of 3 mm and 6 mm, respectively (Fig.~\ref{fig:Agilus_kappa}). To determine angular dissipation $\Gamma$, we place the rubber band on the linkage system, and we apply an angular perturbation to the free edge. We measure the angular displacement and extract $\Gamma$ from the decay rate. We find $\Gamma\approx 7.38\ \mathrm{mN\cdot m \cdot s/rad}$ and $\Gamma\approx 22.04\ \mathrm{mN\cdot m \cdot s/rad}$ for the rubber band with a thickness of 3 mm and 6 mm, respectively (Fig.~\ref{fig:Agilus_Gamma}). We emphasise that these measurements use a Kelvin-Voigt model for the Agilus 30, which is an oversimplification~\cite{dykstraViscoelasticSnappingMetamaterials2019}. However, they indeed demonstrate the dramatically increased dissipation $\Gamma$ of the Agilus relative to a silicone rubber.

\subsection{Unbuckled chain}
We assemble a 22-vertex chain, damped using $\SI{6}{\milli\metre}$ Agilus-30 hinges, loosely confined at either end with masking tape, and not under compression or tension. The non-reciprocity $k^\mathrm{o}=34 \mathrm{mN\cdot m \cdot s/rad}$, and a manual perturbation is applied to the chain's edge, and midpoint [Fig.~2(a,~b)]. 

\subsection{Buckled chain}
For the chain buckling experiments shown in Figs.~\ref{fig:buckling},~\ref{fig:Setup_buckle}, we assemble an open 12-vertex filament, with each hinge damped using the $\SI{6}{\milli\metre}$ Agilus 30. The boundary conditions mimic a clamped-tangent boundary condition in classical buckling: At each end of the chain, the two extremal motors ($i=1,2,11,12$) are clamped in position. The boundary motors $(i=1,12)$ are clamped to external rigid bars via a custom 3D-printed pinch, which is displaced and position controlled to apply the external compression $\text{d} l$. The boundary conditions on the non-reciprocal torques mimic those implemented in our discrete theory \S\ref{sec:TrussTheory}. For $i=1,12$, the active torque $\tau_i=0$. For $i=2$, $\tau_2 = k^\mathrm{o} \delta\theta_3$. For $i=11$, $\tau_{11} = -k^\mathrm{o} \delta\theta_{10}$. 

\subsection{Buckled von Mises truss}
For the 4-bar linkage experiments shown in Fig.~\ref{fig:Truss}, the basic setup is identical to the longer chain. We first buckle the linkage with a strain $\delta l/(3l_0)=0.04$, and then sweep the non-reciprocity in digital units as $k^\mathrm{o}= \SIrange{-0.0035}{0.0035}{\bit}$ in steps of $\SI{0.0002}{\bit}$. Using our calibration data, this sweep is $k^\mathrm{o}= \SIrange{-3}{3}{\milli\newton\metre\per\radian}$ in steps of $\SI{0.17}{\milli\newton\metre\per\radian}$ in physical units. Our protocol starts from $k^\mathrm{o}=0$, sweeps upwards to positive $k^\mathrm{o}$, resets to zero, and then sweeps downwards to negative $k^\mathrm{o}$.
At each $k^\mathrm{o}$ value, we record a movie of the truss vertex positions for $30$ seconds. To calculate the snapping frequencies shown in Fig.~\ref{fig:Truss}(e), we Fourier transform the centre of mass motion of this data, and extract the dominant frequency. 

We show three sets of data in Fig.~\ref{fig:Truss}(e). In the first set, we use the simple silicone rubber of thickness $\SI{3}{\milli\metre}$. To fit our theory, we calculate the location of the CEP, $k^\mathrm{o}_c/B = 1/2\sqrt{5}$ from~\eqref{eq:ThresholdSI}, using our calibrated value of $B$. Next, we predict our rescaled oscillation frequency $\omega \Gamma/B$ by taking the imaginary part of the eigenvalues $\lambda$ of $J$ at the CEP~\eqref{eq:EigenValues}, in the incompressible limit $k\rightarrow \infty$, and using our calibrated value of $\Gamma$. As a check on our results, our torque motors can also implement a digital passive stiffness $B$ as well as the non-reciprocal response $k^\mathrm{o}$. We show two additional datasets in Fig.~3 with two different values of this purely digital stiffness, $B\approx \SIlist{6.8;8.5}{\milli\newton\metre\per\radian}$, and no added rubber restraint. These digital data have the advantage of a clean value of $B$, and hence a clean prediction of the threshold non-reciprocity $k^\mathrm{o}_c/B = 1/2\sqrt{5}$. The disadvantage is that $\Gamma$ is no longer easily measured, because dissipation is now dominated by interactions with the air table. As such, $\Gamma$ for these digital data cannot be independently calibrated, and we simply report a best-fit prediction.

\subsection{Crawling experiments}
To construct the crawler shown in Fig.~\ref{fig:Locomotion}, we construct a brace of the same dimensions as a 6-vertex open chain, which we affix the chain to in a buckled configuration. Each hinge is damped using the 3 mm Agilus 30 rubber band. Our brace is made by laser cutting a rounded rectangle of PMMA of length $\approx6l_0$. Additional slats are cut through the PMMA at either end, which we affix nodes $i=1,2,5,6$ of the chain to via friction. These slats allow us to tuneably compress the chain, and then lock it into position. For the data shown in Fig.~4, $\delta l/l_0 \approx 0.1$. To ensure frictional contact of the crawler with substrate, we tilt the table by $0.6^{\circ}$ to apply $\vec{g}_{\text{eff}}\approx 0.1\ \mathrm{m}\ \mathrm{s}^{-2}$ ($\vec{g}\approx 9.78\ \mathrm{m}\ \mathrm{s}^{-2}$), and attach additional pieces of silicone rubber to the base of each vertex.
\begin{figure}[htbp!]
    \centering
    \includegraphics[width=\linewidth]{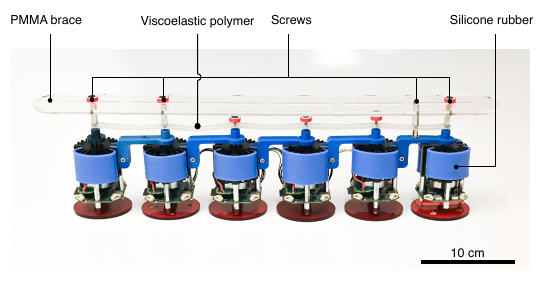}
    \caption{{\bf Details of crawling experiment setup.} A chain of 6 units confined via a laser cutting PMMA brace. By changing the distance between the edge unit and fastening the screws, we can maintain the chain in a buckled state. An Agilus 30 viscoelastic polymer scaffold is attached to the top of the chain to provide hinge elasticity and dissipation. Additional pieces of silicone rubber are put on the side of each unit to ensure frictional contact with the substrate. Scale bar \SI{10}{\centi\metre}.}
    \label{fig:Setup_crawler}
\end{figure}
\begin{figure}[t]
    \centering
    \includegraphics[width=\linewidth]{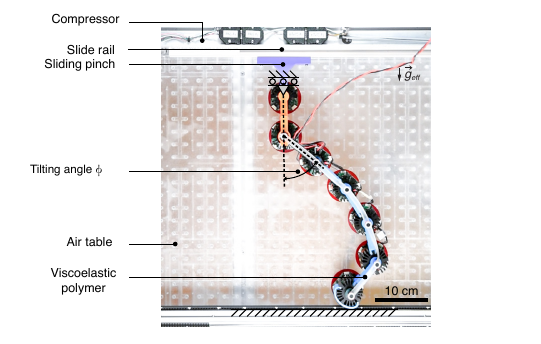}
    \caption{{\bf Details of walking experiment setup.} A chain of 6 units whose the top-end unit is clamped to a slide rail and the bottom-end unit can freely move. The chain is tilted to an angle $\phi$ by fastening the screw on the orange brace. An Agilus 30 viscoelastic polymer scaffold is attached to the top of the chain to provide hinge elasticity and dissipation. Scale bar \SI{10}{\centi\metre}.}
    \label{fig:Setup_walker}
\end{figure}
\begin{figure*}[t!]
    \centering
    \includegraphics[width=\linewidth]{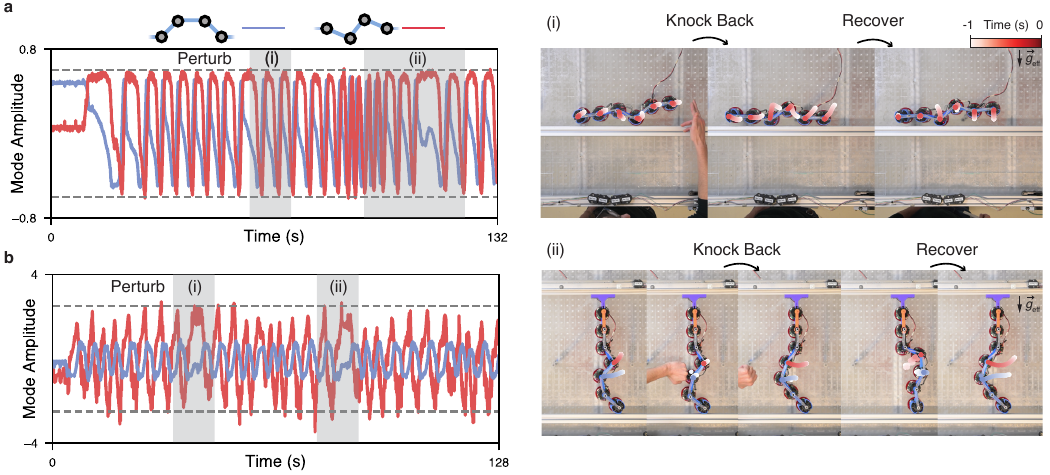}
    \caption{{\bf Crawling and stepping are robust to perturbations.} We manually perturb (a) our crawling chain and (b) the stepping motion that underpins walking and digging. In the face of repeated perturbations from different angles we find that each dynamics is attracted back to the same limit cycle (see also Supplementary Video 5). (a)(i) and (b)(ii) show selected snapshots of this perturbation and recovery process. Symmetric and antisymmetric modes defined as in Fig.~\ref{fig:Locomotion}.}
    \label{fig:Robustness}
\end{figure*}

\subsection{Digging experiments}
In the digging experiments shown in Fig.~\ref{fig:Locomotion}, we cover our air table with a monodisperse mixture of steel bearings of mass $\SI{1.5}{\gram}$ and radius $\SI{0.8}{\milli\metre}$, bound on three sides by the constraints of the table itself. We manually compress a 6-vertex open filament into this granular pile. Each hinge is damped using the 3 mm Agilus 30 rubber band. One end of the filament is clamped to a fixed bar in a manner identical to our buckling experiments, and the other end is free. For these digging experiments, we set the gain $k^\mathrm{o}= \SI{42.3}{\milli\newton\metre}$ to a near-maximal value, such that the motors are effectively saturated at $\tau_{max}$ throughout excavation. We apply a series of compressions manually: we compress the bar a fixed distance, wait until excavation ceases, further compress, and so on.

\subsection{Walking experiments}
In the walking experiments shown in Fig.~\ref{fig:Locomotion}, we again clamp one end of a 6-vertex chain, and leave the other free as in our digging experiments. The key difference is that we now modify the clamp itself such that it can translate horizontally (transverse to the filament), whilst remaining constrained vertically (along the filament). We enforce the tilt angle $\phi$ shown in Fig.~\ref{fig:Locomotion}(h, i) with an additional brace (orange) on the chain itself. Because our focus is on the effects of $\phi$, throughout these experiments we fix a strain $\delta l/5l_{0} = 0.23$ and $k^\mathrm{o} = \SI{8.46}{\milli\newton\metre}$---as in the digging experiments, this $k^\mathrm{o}$ is large enough that the motors saturate at $\tau_{max}$ during the walking stroke. To ensure frictional contact of the foot with substrate during walking, we tilt the table by $2.8^{\circ}$ to apply $\vec{g}_{\text{eff}}\approx 0.5\ \mathrm{m}\ \mathrm{s}^{-2}$ ($\vec{g}\approx 9.78\ \mathrm{m}\ \mathrm{s}^{-2}$), and attach an additional piece of silicone rubber to the base of the foot.

\subsection{Additional Robustness Tests}
To further check the robustness of crawling, digging and walking (Fig.~\ref{fig:Locomotion}), we manually perturb both the limit cycle of our crawler, and the limit cycle for the fundamental stepping motion which underpins walking and digging (Fig.~\ref{fig:Robustness}, Supplementary Video 5). We observe that the dynamics is repeatedly attracted back to the same limit cycle in the face of multiple perturbations.

\let\oldaddcontentsline\addcontentsline
\renewcommand{\addcontentsline}[3]{}
\section*{Acknowledgements}
All the codes and data supporting this study are available on the public repository at 10.5281/zenodo.17513595. S.C.A.-I. and Y.D contributed equally to this work. S.C.A.-I., Y.D., J.V., A.S., A.C., and C.C. and J.B. designed research; S.C.A.-I., Y.D., J.V., R.G.M., A.C., C.C., and J.B. performed research; S.C.A.-I., Y.D., J.V., and J.B. analyzed data; R.G.M., A.S., and A.C. feedback on paper; and S.C.A.-I., C.C., and Y.D. and J.B. wrote the paper. We thank D. Giesen, R. Hassing, S. Koot, E. Hop, and T. van Klingeren for technical assistance. S.C.A.-I. was partially supported by funding from the European Union’s Horizon Europe research and innovation programme under the Marie Skłodowska-Curie postdoctoral fellowship No. 101106384 and the Australian Research Council Centre of Excellence for the Mathematical Modelling of Cellular Systems (MACSYS, CE230100001). J.B. acknowledges funding from the European Union’s Horizon research and  innovation programme under the Marie Sklodowska-Curie Grant Agreement No. 101106500. Y.D. acknowledges financial support from the China Scholarship Council. A.C. acknowledges funding from the Research Council of Norway, project No. 39341989 and from the European Research Council under grant agreement ERC-CoG 101169717. C.C. acknowledges funding from the Netherlands Organization for Scientific Research under grant agreement VI.Vidi.213.131 and from the European Research Council under grant agreement ERC-CoG 101170693.

\bibliography{subsetBiblio}

@article{tongDiscreteDifferentialGeometry2026,
  title = {Discrete Differential Geometry for Simulating Nonlinear Behaviors of Flexible Systems: {{A}} Survey},
  shorttitle = {Discrete Differential Geometry for Simulating Nonlinear Behaviors of Flexible Systems},
  author = {Tong, Dezhong and Choi, Andrew and Wang, Jiaqi and Huang, Weicheng and Chen, Zexiong and Li, Jiahao and Huang, Xiaonan and Liu, Mingchao and Gao, Huajian and Hsia, K. Jimmy},
  year = 2026,
  month = jan,
  journal = {Extreme Mechanics Letters},
  volume = {82},
  pages = {102430},
  issn = {2352-4316},
  doi = {10.1016/j.eml.2025.102430},
  urldate = {2026-01-09},
}

@Article{huangTutorialSimulatingNonlinear2025,
  title = {A Tutorial on Simulating Nonlinear Behaviors of Flexible Structures with the Discrete Differential Geometry ({{DDG}}) Method},
  author = {Huang, Weicheng and Hao, Zhuonan and Li, Jiahao and Tong, Dezhong and Guo, Kexin and Zhang, Yingchao and Gao, Huajian and Hsia, K. Jimmy and Liu, Mingchao},
  year = {2025},
  journal = {Applied Mechanics Reviews},
  volume={1},
  issn = {0003-6900},
  doi = {10.1115/1.4069025},
  urldate = {2026-01-09},
  pages         = {1–88}
}

@Article{pandeyDynamicsSnappingBeams2014,
  title = {Dynamics of Snapping Beams and Jumping Poppers},
  author = {Pandey, A. and Moulton, D. E. and Vella, D. and Holmes, D. P.},
  year = 2014,
  month = feb,
  journal = {Europhysics Letters},
  volume = {105},
  number = {2},
  pages = {24001},
  publisher = {{EDP Sciences, IOP Publishing and Societ\`a Italiana di Fisica}},
  issn = {0295-5075},
  doi = {10.1209/0295-5075/105/24001},
  urldate = {2026-01-08},
  langid = {english},
}

@Article{radissonElasticSnapthroughInstabilities2023,
  title = {Elastic Snap-through Instabilities Are Governed by Geometric Symmetries},
  author = {Radisson, Basile and Kanso, Eva},
  year = 2023,
  month = jun,
  journal = {Physical Review Letters},
  volume = {130},
  number = {23},
  pages = {236102},
  issn = {0031-9007, 1079-7114},
  doi = {10.1103/PhysRevLett.130.236102},
  urldate = {2026-01-08},
}

@Article{gomezCriticalSlowingPurely2017,
  title = {Critical Slowing down in Purely Elastic `Snap-through' Instabilities},
  author = {Gomez, Michael and Moulton, Derek E. and Vella, Dominic},
  year = 2017,
  month = feb,
  journal = {Nature Physics},
  volume = {13},
  number = {2},
  pages = {142--145},
  publisher = {Nature Publishing Group},
  issn = {1745-2481},
  doi = {10.1038/nphys3915},
  urldate = {2026-01-08},
  copyright = {2016 Springer Nature Limited},
  langid = {english},
}

@Article{liuDelayedBifurcationElastic2021,
  title = {Delayed Bifurcation in Elastic Snap-through Instabilities},
  author = {Liu, Mingchao and Gomez, Michael and Vella, Dominic},
  year = 2021,
  month = jun,
  journal = {Journal of the Mechanics and Physics of Solids},
  volume = {151},
  pages = {104386},
  issn = {00225096},
  doi = {10.1016/j.jmps.2021.104386},
  urldate = {2026-01-05},
  langid = {english},
}

@misc{veenstraWaveCoarseningDrives2025,
  title = {Wave Coarsening Drives Time Crystallization in Active Solids},
  author = {Veenstra, Jonas and Binysh, Jack and Seinen, Vito and Naber, Rutger and {Robledo-Poisson}, Damien and Hunt, Andres and van Saarloos, Wim and Souslov, Anton and Coulais, Corentin},
  year = 2025,
  month = aug,
  number = {arXiv:2508.20052},
  eprint = {2508.20052},
  publisher = {arXiv},
  doi = {10.48550/arXiv.2508.20052},
  urldate = {2025-09-01},
  archiveprefix = {arXiv},
}

@Article{zellaUniversal2024,
  title = {Universal Phenomenology at Critical Exceptional Points of Nonequilibrium $\mathrm{O}(N)$ Models},
  author = {Zelle, Carl Philipp and Daviet, Romain and Rosch, Achim and Diehl, Sebastian},
  journal = {Phys. Rev. X},
  volume = {14},
  issue = {2},
  pages = {021052},
  numpages = {44},
  year = {2024},
  month = {Jun},
  publisher = {American Physical Society},
  doi = {10.1103/PhysRevX.14.021052},
  url = {https://link.aps.org/doi/10.1103/PhysRevX.14.021052}
}

@Article{guoLimblessUndulatoryPropulsion2008,
  title = {Limbless Undulatory Propulsion on Land},
  author = {Guo, Z. V. and Mahadevan, L.},
  year = {2008},
  month = mar,
  journal = {Proceedings of the National Academy of Sciences},
  volume = {105},
  number = {9},
  pages = {3179--318}
}

@Article{         siefertBioinspiredPneumaticShapemorphing2019,
  title         = {Bio-Inspired Pneumatic Shape-Morphing Elastomers},
  author        = {Si{\'e}fert, Emmanuel and Reyssat, Etienne and Bico,
                  Jos{\'e} and Roman, Beno{\^i}t},
  year          = {2019},
  month         = jan,
  journal       = {Nature Materials},
  volume        = {18},
  number        = {1},
  pages         = {24--28},
  publisher     = {Nature Publishing Group},
  issn          = {1476-4660},
  doi           = {10.1038/s41563-018-0219-x},
  urldate       = {2025-10-16},
  copyright     = {2018 The Author(s), under exclusive licence to Springer
                  Nature Limited},
  langid        = {english}
}

@Article{         cammannFormFunctionBiological2025a,
  title         = {Form and Function in Biological Filaments: A Physicist's
                  Review},
  shorttitle    = {Form and Function in Biological Filaments},
  author        = {Cammann, Jan and {Laeverenz-Schlogelhofer}, Hannah and
                  Wan, Kirsty Y. and Mazza, Marco G.},
  year          = {2025},
  month         = sep,
  journal       = {Philosophical Transactions of the Royal Society A:
                  Mathematical, Physical and Engineering Sciences},
  volume        = {383},
  number        = {2304},
  pages         = {20240253},
  publisher     = {Royal Society},
  doi           = {10.1098/rsta.2024.0253},
  urldate       = {2025-10-07}
}

@Article{         jiangEnergyPrinciplesNonHermitian2025,
  title         = {Energy Principles for Non-{{Hermitian}} Active Metabeams
                  with Odd Elasticity},
  author        = {Jiang, Jiaqing and Chen, Weiqiu and Amabili, Marco},
  year          = {2025},
  month         = oct,
  journal       = {International Journal of Mechanical Sciences},
  volume        = {304},
  pages         = {110692},
  issn          = {00207403},
  doi           = {10.1016/j.ijmecsci.2025.110692},
  urldate       = {2025-10-02},
  langid        = {english}
}

@Misc{            nemethNonreciprocalConstitutiveLaws2025,
  title         = {Nonreciprocal Constitutive Laws for Oriented Active
                  Solids},
  author        = {N{\'e}meth, Bal{\'a}zs and Kobayashi, Takuya and Adhikari,
                  Ronojoy},
  year          = {2025},
  month         = sep,
  number        = {arXiv:2509.11430},
  eprint        = {2509.11430},
  publisher     = {arXiv},
  doi           = {10.48550/arXiv.2509.11430},
  urldate       = {2025-09-16},
  archiveprefix = {arXiv}
}

@Article{sinaasappelParticleSweepingCollection2026,
  title = {Particle {{Sweeping}} and {{Collection}} by {{Active}} and {{Living Filaments}}},
  author = {Sinaasappel, R. and Prathyusha, K. R. and Tuazon, H. and Mirzahossein, E. and Illien, P. and Bhamla, S. and Deblais, A.},
  year = 2026,
  month = jan,
  journal = {Physical Review X},
  volume = {16},
  number = {1},
  pages = {011003},
  publisher = {American Physical Society},
  doi = {10.1103/yxp1-t43g},
  urldate = {2026-02-24},
}

@Article{         dudekShapeMorphingMetamaterials2025,
  title         = {Shape-Morphing Metamaterials},
  author        = {Dudek, Krzysztof K. and Kadic, Muamer and Coulais,
                  Corentin and Bertoldi, Katia},
  year          = {2025},
  month         = oct,
  journal       = {Nature Reviews Materials},
  volume        = {10},
  number        = {10},
  pages         = {783--798},
  publisher     = {Nature Publishing Group},
  issn          = {2058-8437},
  doi           = {10.1038/s41578-025-00828-9},
  urldate       = {2025-10-02},
  copyright     = {2025 Springer Nature Limited},
  langid        = {english}
}

@Article{         brennerOptimalDesignBistable2003,
  title         = {Optimal Design of a Bistable Switch},
  author        = {Brenner, Michael P. and Lang, Jeffrey H. and Li, Jian and
                  Qiu, Jin and Slocum, Alexander H.},
  year          = {2003},
  month         = aug,
  journal       = {Proceedings of the National Academy of Sciences},
  volume        = {100},
  number        = {17},
  pages         = {9663--9667},
  issn          = {0027-8424, 1091-6490},
  doi           = {10.1073/pnas.1531507100},
  urldate       = {2025-10-02},
  langid        = {english}
}

@Article{         dykstraBucklingMetamaterialsExtreme2023,
  title         = {Buckling {{Metamaterials}} for {{Extreme Vibration
                  Damping}}},
  author        = {Dykstra, David M.J. and Lenting, Coen and Masurier,
                  Alexandre and Coulais, Corentin},
  year          = {2023},
  journal       = {Advanced Materials},
  volume        = {35},
  number        = {35},
  pages         = {2301747},
  issn          = {1521-4095},
  doi           = {10.1002/adma.202301747},
  urldate       = {2025-10-02},
  langid        = {english}
}

@Article{         wangInsectscaleJumpingRobots2023,
  title         = {Insect-Scale Jumping Robots Enabled by a Dynamic Buckling
                  Cascade},
  author        = {Wang, Yuzhe and Wang, Qiong and Liu, Mingchao and Qin,
                  Yimeng and Cheng, Liuyang and Bolmin, Ophelia and Alleyne,
                  Marianne and Wissa, Aimy and Baughman, Ray H. and Vella,
                  Dominic and Tawfick, Sameh},
  year          = {2023},
  month         = jan,
  journal       = {Proceedings of the National Academy of Sciences},
  volume        = {120},
  number        = {5},
  pages         = {e2210651120},
  publisher     = {Proceedings of the National Academy of Sciences},
  doi           = {10.1073/pnas.2210651120},
  urldate       = {2025-08-27}
}

@Article{         forterreHowVenusFlytrap2005,
  title         = {How the {{Venus}} Flytrap Snaps},
  author        = {Forterre, Yo{\"e}l and Skotheim, Jan M. and Dumais,
                  Jacques and Mahadevan, L.},
  year          = {2005},
  month         = jan,
  journal       = {Nature},
  volume        = {433},
  number        = {7024},
  pages         = {421--425},
  publisher     = {Nature Publishing Group},
  issn          = {1476-4687},
  doi           = {10.1038/nature03185},
  urldate       = {2025-10-02},
  copyright     = {2004 Macmillan Magazines Ltd.},
  langid        = {english}
}

@Article{         sonBacteriaCanExploit2013,
  title         = {Bacteria Can Exploit a Flagellar Buckling Instability to
                  Change Direction},
  author        = {Son, Kwangmin and Guasto, Jeffrey S. and Stocker, Roman},
  year          = {2013},
  journal  = {Nature Physics},
  shortjournal  = {Nature Phys},
  volume        = {9},
  number        = {8},
  pages         = {494--498},
  publisher     = {Nature Publishing Group},
  issn          = {1745-2481},
  doi           = {10.1038/nphys2676},
  urldate       = {2025-10-02},
  langid        = {english}
}

@Article{         dykstraViscoelasticSnappingMetamaterials2019,
  title         = {Viscoelastic {{Snapping Metamaterials}}},
  author        = {Dykstra, David M. J. and Busink, Joris and Ennis, Bernard
                  and Coulais, Corentin},
  year          = {2019},
  month         = sep,
  journal       = {Journal of Applied Mechanics},
  volume        = {86},
  pages        = {111012},
  issn          = {0021-8936},
  urldate       = {2025-09-01}
}

@Article{         Buckmaster_Nachman_Ting_1975,
  title         = {The buckling and stretching of a viscida},
  volume        = {69},
  doi           = {10.1017/S0022112075001279},
  number        = {1},
  journal       = {Journal of Fluid Mechanics},
  author        = {Buckmaster, J. D. and Nachman, A. and Ting, L.},
  year          = {1975},
  pages         = {1–20}
}

@PhDThesis{       kodio_thesis,
  author        = {Kodio, Ousmane},
  title         = {Dynamic Buckling Instabilities in Fluids and Solids},
  school        = {University of Oxford},
  year          = {2019}
}

@Article{         avniDynamicalPhaseTransitions2025,
  title         = {Dynamical Phase Transitions in the Nonreciprocal {{Ising}}
                  Model},
  author        = {Avni, Yael and Fruchart, Michel and Martin, David and
                  Seara, Daniel and Vitelli, Vincenzo},
  year          = {2025},
  month         = mar,
  journal       = {Physical Review E},
  volume        = {111},
  number        = {3},
  pages         = {034124},
  publisher     = {American Physical Society},
  doi           = {10.1103/PhysRevE.111.034124},
  urldate       = {2025-04-01}
}

@Article{         brackenburyCaterpillarKinematics1997,
  title         = {Caterpillar Kinematics},
  author        = {Brackenbury, John},
  year          = {1997},
  month         = dec,
  journal       = {Nature},
  volume        = {390},
  number        = {6659},
  pages         = {453--453},
  publisher     = {Nature Publishing Group},
  issn          = {1476-4687},
  doi           = {10.1038/37253},
  urldate       = {2025-06-18},
  copyright     = {1997 Macmillan Magazines Ltd.},
  langid        = {english}
}

@Article{         braunsNonreciprocalPatternFormation2024,
  title         = {Nonreciprocal {{Pattern Formation}} of {{Conserved
                  Fields}}},
  author        = {Brauns, Fridtjof and Marchetti, M. Cristina},
  year          = {2024},
  month         = apr,
  journal       = {Physical Review X},
  volume        = {14},
  number        = {2},
  pages         = {021014},
  issn          = {2160-3308},
  doi           = {10.1103/PhysRevX.14.021014},
  urldate       = {2024-12-12},
  langid        = {english}
}

@Article{         cassReactiondiffusionBasisAnimated2023,
  title         = {The Reaction-Diffusion Basis of Animated Patterns in
                  Eukaryotic Flagella},
  author        = {Cass, James F. and {Bloomfield-Gad{\^e}lha}, Hermes},
  year          = {2023},
  month         = sep,
  journal       = {Nature Communications},
  volume        = {14},
  number        = {1},
  pages         = {5638},
  issn          = {2041-1723},
  doi           = {10.1038/s41467-023-40338-2},
  urldate       = {2025-06-28},
  langid        = {english}
}

@Article{         chengOddElasticityRealized2021,
  title         = {Odd Elasticity Realized by Piezoelectric Material with
                  Linear Feedback},
  author        = {Cheng, Wen and Hu, Gengkai},
  year          = {2021},
  month         = sep,
  journal       = {Science China Physics, Mechanics \& Astronomy},
  volume        = {64},
  number        = {11},
  pages         = {2},
  issn          = {1869-1927},
  doi           = {10.1007/s11433-021-1756-0},
  urldate       = {2024-10-02},
  langid        = {english}
}

@Article{         chenRealizationActiveMetamaterials2021,
  title         = {Realization of Active Metamaterials with Odd Micropolar
                  Elasticity},
  author        = {Chen, Yangyang and Li, Xiaopeng and Scheibner, Colin and
                  Vitelli, Vincenzo and Huang, Guoliang},
  year          = {2021},
  month         = oct,
  journal       = {Nature Communications},
  volume        = {12},
  number        = {1},
  pages         = {5935},
  publisher     = {Nature Publishing Group},
  issn          = {2041-1723},
  doi           = {10.1038/s41467-021-26034-z},
  urldate       = {2022-08-17},
  copyright     = {2021 The Author(s)},
  langid        = {english}
}

@Article{         clarkeBifurcationsNonlinearDynamics2024,
  title         = {Bifurcations and Nonlinear Dynamics of the Follower Force
                  Model for Active Filaments},
  author        = {Clarke, Bethany and Hwang, Yongyun and Keaveny, Eric E.},
  year          = {2024},
  month         = jul,
  journal       = {Physical Review Fluids},
  volume        = {9},
  number        = {7},
  pages         = {073101},
  issn          = {2469-990X},
  doi           = {10.1103/PhysRevFluids.9.073101},
  urldate       = {2025-05-18},
  langid        = {english}
}

@Article{         decanioSpontaneousOscillationsElastic2017,
  title         = {Spontaneous Oscillations of Elastic Filaments Induced by
                  Molecular Motors},
  author        = {De Canio, Gabriele and Lauga, Eric and Goldstein, Raymond
                  E.},
  year          = {2017},
  month         = nov,
  journal       = {Journal of The Royal Society Interface},
  volume        = {14},
  number        = {136},
  pages         = {20170491},
  issn          = {1742-5689, 1742-5662},
  doi           = {10.1098/rsif.2017.0491},
  urldate       = {2024-07-25},
  langid        = {english}
}

@Article{         dengNonlinearWavesFlexible2021,
  title         = {Nonlinear Waves in Flexible Mechanical Metamaterials},
  author        = {Deng, B. and Raney, J. R. and Bertoldi, K. and Tournat,
                  V.},
  year          = {2021},
  month         = jul,
  journal       = {Journal of Applied Physics},
  volume        = {130},
  number        = {4},
  pages         = {040901},
  issn          = {0021-8979, 1089-7550},
  doi           = {10.1063/5.0050271},
  urldate       = {2025-05-18},
  langid        = {english}
}

@Article{Doi2011,
  title = {Onsager's Variational Principle in Soft Matter},
  author = {Doi, Masao},
  year = 2011,
  month = jun,
  journal = {Journal of Physics: Condensed Matter},
  volume = {23},
  number = {28},
  pages = {284118},
  issn = {0953-8984},
  doi = {10.1088/0953-8984/23/28/284118},
  urldate = {2025-10-16},
  langid = {english},
}

@Book{            doiSoftMatterPhysics2013,
  title         = {Soft Matter Physics},
  author        = {Doi, M.},
  year          = {2013},
  publisher     = {Oxford University Press},
  isbn          = {978-0-19-965295-2},
  langid        = {english},
  lccn          = {QC173.458.S62 D65 2013}
}

@Article{         filyBucklingInstabilitiesSpatiotemporal2020,
  title         = {Buckling Instabilities and Spatio-Temporal Dynamics of
                  Active Elastic Filaments},
  author        = {Fily, Yaouen and Subramanian, Priya and Schneider, Tobias
                  M. and Chelakkot, Raghunath and Gopinath, Arvind},
  year          = {2020},
  month         = apr,
  journal       = {Journal of The Royal Society Interface},
  volume        = {17},
  number        = {165},
  pages         = {20190794},
  publisher     = {Royal Society},
  doi           = {10.1098/rsif.2019.0794},
  urldate       = {2025-05-19}
}

@Article{         fossatiOddElasticityTopological2024,
  title         = {Odd Elasticity and Topological Waves in Active Surfaces},
  author        = {Fossati, Michele and Scheibner, Colin and Fruchart, Michel
                  and Vitelli, Vincenzo},
  year          = {2024},
  month         = feb,
  journal       = {Physical Review E},
  volume        = {109},
  number        = {2},
  pages         = {024608},
  issn          = {2470-0045, 2470-0053},
  doi           = {10.1103/PhysRevE.109.024608},
  urldate       = {2025-01-16},
  langid        = {english}
}

@Article{         fruchartNonreciprocalPhaseTransitions2021,
  title         = {Non-Reciprocal Phase Transitions},
  author        = {Fruchart, Michel and Hanai, Ryo and Littlewood, Peter B.
                  and Vitelli, Vincenzo},
  year          = {2021},
  month         = apr,
  journal       = {Nature},
  volume        = {592},
  number        = {7854},
  pages         = {363--369},
  publisher     = {Nature Publishing Group},
  issn          = {1476-4687},
  doi           = {10.1038/s41586-021-03375-9},
  urldate       = {2025-04-01},
  copyright     = {2021 The Author(s), under exclusive licence to Springer
                  Nature Limited},
  langid        = {english}
}

@InCollection{    golubitskyBurstingCoupledCell2005,
  title         = {Bursting in Coupled Cell Systems},
  booktitle     = {Bursting},
  author        = {Golubitsky, Martin and Josi\'c, Kre\^simir and Shiau,
                  LieJune},
  year          = {2005},
  month         = oct,
  pages         = {201--221},
  publisher     = {World Scientific},
  doi           = {10.1142/9789812703231_0008},
  urldate       = {2025-06-12},
  isbn          = {978-981-256-506-8}
}

@Article{         hanaiCriticalFluctuationsManybody2020,
  title         = {Critical Fluctuations at a Many-Body Exceptional Point},
  author        = {Hanai, Ryo and Littlewood, Peter B.},
  year          = {2020},
  month         = jul,
  journal       = {Physical Review Research},
  volume        = {2},
  number        = {3},
  pages         = {033018},
  issn          = {2643-1564},
  doi           = {10.1103/PhysRevResearch.2.033018},
  urldate       = {2025-05-08},
  langid        = {english}
}

@Article{         hanaiNonreciprocalFrustrationTime2024,
  title         = {Nonreciprocal {{Frustration}}: {{Time Crystalline
                  Order-by-Disorder Phenomenon}} and a {{Spin-Glass-like
                  State}}},
  shorttitle    = {Nonreciprocal {{Frustration}}},
  author        = {Hanai, Ryo},
  year          = {2024},
  month         = feb,
  journal       = {Physical Review X},
  volume        = {14},
  number        = {1},
  pages         = {011029},
  publisher     = {American Physical Society},
  doi           = {10.1103/PhysRevX.14.011029},
  urldate       = {2025-01-18}
}

@Article{         heydariSeaStarInspired2020,
  title         = {Sea Star Inspired Crawling and Bouncing},
  author        = {Heydari, Sina and Johnson, Amy and Ellers, Olaf and
                  McHenry, Matthew J. and Kanso, Eva},
  year          = {2020},
  month         = jan,
  journal       = {Journal of The Royal Society Interface},
  volume        = {17},
  number        = {162},
  pages         = {20190700},
  publisher     = {Royal Society},
  doi           = {10.1098/rsif.2019.0700},
  urldate       = {2025-06-09}
}

@Book{            howellCompliantMechanisms,
  title         = {Compliant {{Mechanisms}}},
  author        = {Howell, Larry L},
  langid        = {english},
  year          = {2001},
  publisher     = {Wiley}
}

@Article{         huMechanicsSlitheringLocomotion2009,
  title         = {The Mechanics of Slithering Locomotion},
  author        = {Hu, David L. and Nirody, Jasmine and Scott, Terri and
                  Shelley, Michael J.},
  year          = {2009},
  month         = jun,
  journal       = {Proceedings of the National Academy of Sciences},
  volume        = {106},
  number        = {25},
  pages         = {10081--10085},
  issn          = {0027-8424, 1091-6490},
  doi           = {10.1073/pnas.0812533106},
  urldate       = {2025-05-19},
  langid        = {english}
}

@Article{         ishimotoSelforganizedSwimmingOdd2022,
  title         = {Self-Organized Swimming with Odd Elasticity},
  author        = {Ishimoto, Kenta and Moreau, Cl{\'e}ment and Yasuda, Kento},
  year          = {2022},
  month         = jun,
  journal       = {Physical Review E},
  volume        = {105},
  number        = {6},
  pages         = {064603},
  publisher     = {American Physical Society},
  doi           = {10.1103/PhysRevE.105.064603},
  urldate       = {2025-06-10}
}

@Article{         kwakernaakCountingSequentialInformation2023,
  title         = {Counting and {{Sequential Information Processing}} in
                  {{Mechanical Metamaterials}}},
  author        = {Kwakernaak, Lennard J. and {van Hecke}, Martin},
  year          = {2023},
  month         = jun,
  journal       = {Physical Review Letters},
  volume        = {130},
  number        = {26},
  pages         = {268204},
  publisher     = {American Physical Society},
  doi           = {10.1103/PhysRevLett.130.268204},
  urldate       = {2025-07-19}
}

@Article{         leeMechanicalNeuralNetworks2022,
  title         = {Mechanical Neural Networks: {{Architected}} Materials That
                  Learn Behaviors},
  shorttitle    = {Mechanical Neural Networks},
  author        = {Lee, Ryan H. and Mulder, Erwin A. B. and Hopkins, Jonathan
                  B.},
  year          = {2022},
  month         = oct,
  journal       = {Science Robotics},
  volume        = {7},
  number        = {71},
  pages         = {eabq7278},
  issn          = {2470-9476},
  doi           = {10.1126/scirobotics.abq7278},
  urldate       = {2025-05-12},
  langid        = {english}
}

@Article{         melanconInflatableOrigamiMultimodal2022,
  title         = {Inflatable {{Origami}}: {{Multimodal Deformation}} via
                  {{Multistability}}},
  shorttitle    = {Inflatable {{Origami}}},
  author        = {Melancon, David and Forte, Antonio Elia and Kamp, Leon M.
                  and Gorissen, Benjamin and Bertoldi, Katia},
  year          = {2022},
  month         = aug,
  journal       = {Advanced Functional Materials},
  volume        = {32},
  pages         ={2201891}
}

@Article{         miriExceptionalPointsOptics2019,
  title         = {Exceptional Points in Optics and Photonics},
  author        = {Miri, Mohammad-Ali and Al{\`u}, Andrea},
  year          = {2019},
  month         = jan,
  journal       = {Science},
  volume        = {363},
  number        = {6422},
  pages         = {eaar7709},
  issn          = {0036-8075, 1095-9203},
  doi           = {10.1126/science.aar7709},
  urldate       = {2025-01-09},
  langid        = {english}
}

@Article{         nadkarniUnidirectionalTransitionWaves2016,
  title         = {Unidirectional {{Transition Waves}} in {{Bistable
                  Lattices}}},
  author        = {Nadkarni, Neel and Arrieta, Andres F. and Chong,
                  Christopher and Kochmann, Dennis M. and Daraio, Chiara},
  year          = {2016},
  month         = jun,
  journal       = {Physical Review Letters},
  volume        = {116},
  number        = {24},
  pages         = {244501},
  issn          = {0031-9007, 1079-7114},
  doi           = {10.1103/PhysRevLett.116.244501},
  urldate       = {2021-07-20},
  langid        = {english}
}

@Article{         patilUltrafastReversibleSelfassembly2023,
  title         = {Ultrafast Reversible Self-Assembly of Living Tangled
                  Matter},
  author        = {Patil, Vishal P. and Tuazon, Harry and Kaufman, Emily and
                  Chakrabortty, Tuhin and Qin, David and Dunkel, J{\"o}rn and
                  Bhamla, M. Saad},
  year          = {2023},
  month         = apr,
  journal       = {Science},
  volume        = {380},
  number        = {6643},
  pages         = {392--398},
  publisher     = {American Association for the Advancement of Science},
  doi           = {10.1126/science.ade7759},
  urldate       = {2024-02-22}
}

@Article{         purcellLifeLowReynolds1977a,
  title         = {Life at Low {{Reynolds}} Number},
  author        = {Purcell, E. M.},
  year          = {1977},
  month         = jan,
  journal       = {American Journal of Physics},
  volume        = {45},
  number        = {1},
  pages         = {3--11},
  issn          = {0002-9505},
  doi           = {10.1119/1.10903},
  urldate       = {2025-07-18}
}

@Article{         sekimotoSymmetryBreakingInstabilities1995,
  title         = {Symmetry {{Breaking Instabilities}} of an {{In Vitro
                  Biological System}}},
  author        = {Sekimoto, Ken and Mori, Naoki and Tawada, Katsuhisa and
                  Toyoshima, Yoko Y.},
  year          = {1995},
  month         = jul,
  journal       = {Physical Review Letters},
  volume        = {75},
  number        = {1},
  pages         = {172--175},
  publisher     = {American Physical Society},
  doi           = {10.1103/PhysRevLett.75.172},
  urldate       = {2025-06-11}
}

@Article{         suchanekEntropyProductionNonreciprocal2023,
  title         = {Entropy Production in the Nonreciprocal {{Cahn-Hilliard}}
                  Model},
  author        = {Suchanek, Thomas and Kroy, Klaus and Loos, Sarah A. M.},
  year          = {2023},
  month         = dec,
  journal       = {Physical Review E},
  volume        = {108},
  number        = {6},
  pages         = {064610},
  publisher     = {American Physical Society},
  doi           = {10.1103/PhysRevE.108.064610},
  urldate       = {2025-01-09}
}

@Article{         veenstraAdaptiveLocomotionActive2025,
  title         = {Adaptive Locomotion of Active Solids},
  author        = {Veenstra, Jonas and Scheibner, Colin and Brandenbourger,
                  Martin and Binysh, Jack and Souslov, Anton and Vitelli,
                  Vincenzo and Coulais, Corentin},
  year          = {2025},
  month         = mar,
  journal       = {Nature},
  volume        = {639},
  number        = {8056},
  pages         = {935--941},
  publisher     = {Nature Publishing Group},
  issn          = {1476-4687},
  doi           = {10.1038/s41586-025-08646-3},
  urldate       = {2025-10-16},
  copyright     = {2025 The Author(s), under exclusive licence to Springer
                  Nature Limited},
  langid        = {english}
}

@Article{         wangMechanicalIntelligenceSimplifies2023,
  title         = {Mechanical Intelligence Simplifies Control in Terrestrial
                  Limbless Locomotion},
  author        = {Wang, Tianyu and Pierce, Christopher and Kojouharov, Velin
                  and Chong, Baxi and Diaz, Kelimar and Lu, Hang and Goldman,
                  Daniel I.},
  year          = {2023},
  month         = dec,
  journal       = {Science Robotics},
  volume        = {8},
  number        = {85},
  pages         = {eadi2243},
  issn          = {2470-9476},
  doi           = {10.1126/scirobotics.adi2243},
  urldate       = {2025-05-18},
  langid        = {english}
}

@Article{         wangTransientAmplificationBroken2024,
  title         = {Transient {{Amplification}} of {{Broken Symmetry}} in
                  {{Elastic Snap-Through}}},
  author        = {Wang, Qiong and Giudici, Andrea and Huang, Weicheng and
                  Wang, Yuzhe and Liu, Mingchao and Tawfick, Sameh and Vella,
                  Dominic},
  year          = {2024},
  month         = jun,
  journal       = {Physical Review Letters},
  volume        = {132},
  number        = {26},
  pages         = {267201},
  publisher     = {American Physical Society},
  doi           = {10.1103/PhysRevLett.132.267201},
  urldate       = {2025-03-04}
}

@Misc{            weiAutonomousLifelikeBehavior2025,
  title         = {Autonomous Life-like Behavior Emerging in Active and
                  Flexible Microstructures},
  author        = {Wei, Mengshi and Kraft, Daniela J.},
  year          = {2025},
  month         = jun,
  number        = {arXiv:2506.15198},
  eprint        = {2506.15198},
  publisher     = {arXiv},
  doi           = {10.48550/arXiv.2506.15198},
  urldate       = {2025-07-19},
  archiveprefix = {arXiv}
}

@Article{         xiEmergentBehaviorsBucklingdriven2024,
  title         = {Emergent Behaviors of Buckling-Driven Elasto-Active
                  Structures},
  author        = {Xi, Yuchen and Marzin, Tom and Huang, Richard B. and
                  Jones, Trevor J. and Brun, P.-T.},
  year          = {2024},
  month         = nov,
  journal       = {Proceedings of the National Academy of Sciences},
  volume        = {121},
  number        = {45},
  pages         = {e2410654121},
  publisher     = {Proceedings of the National Academy of Sciences},
  doi           = {10.1073/pnas.2410654121},
  urldate       = {2025-10-07}
}

@Article{         zhengSelfOscillationSynchronizationTransitions2023,
  title         = {Self-{{Oscillation}} and {{Synchronization Transitions}}
                  in {{Elastoactive Structures}}},
  author        = {Zheng, Ellen and Brandenbourger, Martin and Robinet, Louis
                  and Schall, Peter and Lerner, Edan and Coulais, Corentin},
  year          = {2023},
  month         = apr,
  journal       = {Physical Review Letters},
  volume        = {130},
  number        = {17},
  pages         = {178202},
  issn          = {0031-9007, 1079-7114},
  doi           = {10.1103/PhysRevLett.130.178202},
  urldate       = {2023-09-25},
  langid        = {english}
}

@Article{active_poly_review2,
author = {Winkler,Roland G.  and Gompper,Gerhard },
title = {The physics of active polymers and filaments},
journal = {The Journal of Chemical Physics},
volume = {153},
number = {4},
pages = {040901},
year = {2020},
doi = {10.1063/5.0011466},
  }

\let\addcontentsline\oldaddcontentsline
\onecolumngrid
\makeatletter 
\def\tagform@#1{\maketag@@@{(S\ignorespaces#1\unskip\@@italiccorr)}}
\makeatother
\graphicspath{{Figures/}} 
\makeatletter
\makeatletter \renewcommand{\fnum@figure}
{\figurename~\thefigure}
\makeatother
\def\eq#1{{Eq.~(S\ref{#1})}}    
\renewcommand{\thefigure}{S\arabic{figure}}       
\setcounter{figure}{0} 
\setcounter{equation}{0} 
\renewcommand{\d}{\partial}
\newcommand{\D}[2]{\frac{\partial #1}{\partial #2}}
\setcounter{secnumdepth}{1} 

\def\thesection{\Roman{section}}
\def\thesubsection{\Roman{section}.\Roman{subsection}}
\def\thesubsubsection{\Roman{section}.\Roman{subsection}.\Roman{subsubsection}}

\title{Supplementary Information: Nonreciprocal buckling makes active filaments polyfunctional}

\maketitle
\onecolumngrid
\section{Introduction}

In this Supplementary Information, we first develop the continuum theory of non-reciprocal filaments in \S\ref{sec:Continuum}. Next we focus 
on the odd von Mises truss. We first provide more details of the $\mathbb{Z}_2$ Bogdanov-Takens bifurcation, which contains the CEP in its linearized dynamics (\S\ref{sec:Takens}). Next we develop microscopic simulations (\S\ref{sec:DiscreteSimulations}) that complement the theory described in the Main Text and capture the bifurcation scenario underpinning non-reciprocal buckling. Finally, we show that a critical exceptional point also underpins the buckling of an entirely different class of active beam composed of follower forces (\S\ref{sec:Follower}).

\section{Continuum theory}\label{sec:Continuum}
\begin{table}[b!]
    \begin{center}
      \begin{tabular}{ccc} 
        \textbf{Name} & \textbf{Symbol} & \textbf{Units}\\ 
        \hline
        \hline
        System Size & $l$ & $L$ \\
        \hline
        \textbf{Microscopic} & & \\
        \hline
        Mass & $m$ & $M$  \\
        Lattice Spacing & $l_0$ & $L$ \\
        Bending Rigidity & $B$ & $E$ \\
        Longitudinal Spring Stiffness & $k$ & $E L^{-2}$ \\
        Microscopic Non-reciprocity & $k^\text{o}$ & $E$ \\
        Hinge Dissipation & $\Gamma$ & $E T$ \\
        \hline
        \textbf{Continuum} & & \\
        \hline
        Line Density & $\rho$ & $E L^{-1}$ \\
        Bending Modulus & $A$ & $E L$ \\
        Continuum Non-reciprocity & $\zeta$ & $E L^2$ \\
        Bending Viscosity & $\eta$ & $E L T$ \\
        \hline
      \end{tabular}
    \end{center}
    \caption{Names, symbols, and units of material parameters.} 
    \label{tab:scales}
\end{table}
In this section we derive a continuum theory for odd, inextensible filaments using a symmetry based approach.
\subsection{Geometry}
We begin by restricting ourselves to $2$D, and defining the midline of our filament by the vector ${\bf x}(s)$. 
The rod has a locally induced frame given by
\begin{equation}
    {\bf t} = \partial_s {\bf x}\text{,} \quad {\bf n} = {\bf J }\cdot {\bf t}\text{,}
\end{equation}
where ${\bf J}$ is an anticlockwise rotation of $\pi/2$ around the axis perpendicular to the plane, and $s$ is the arclength parameter along the curve. 

The geodesic curvature, $\kappa$, is then defined by
\begin{equation}
\begin{split}
    &\partial_s {\bf t} = -\kappa {\bf n}\text{,} \\
    &\partial_s {\bf n} = \kappa {\bf t}\text{.}
\end{split}
\end{equation}

The filament moves with a velocity in the plane given by
\begin{equation}
\dot{\bf x} = W{\bf t} + U {\bf n}\text{,}
\end{equation}
with $W$ the tangential, and $U$ the normal velocity. We will assume that the filament is inextensible, a condition given by
\begin{equation}\label{eq:inextensible}
    \nabla \cdot \dot{\bf x}=\partial_s W + U \kappa = 0\text{,}
\end{equation}
where $\nabla = {\bf t} \partial_s$.

We can now calculate time derivatives of the geometric quantities, namely the tangent vector, $\bf t$, normal vector $\bf n$, and the curvature, $\kappa$. Taking the time derivative of our tangent vector gives
\begin{align}
    \partial_t {\bf t} =  \partial_t \partial_s {\bf x} = \partial_s\partial_t {\bf x} = \partial_s\left( W {\bf t} + U {\bf n} \right) = \left( \partial_s W + U \kappa\right) {\bf t} + \left( \partial_s U - W \kappa\right) {\bf n} = \left( \partial_s U - W \kappa\right) {\bf n}\text{,}
\end{align}
where the last step has used the inextensibility condition,~\eqref{eq:inextensible}. Similarly, for the change in normal vector and curvature:
\begin{align}
    &\partial_t {\bf n} = \left( W \kappa - \partial_s U\right) {\bf t}\text{,} \\
    &\partial_t \kappa =  \partial_t ( - {\bf n}\cdot\partial_s {\bf t})= \partial_s (\kappa W - \partial_s U)\text{.}
\end{align}
With these geometric and kinematic quantities in hand, we can now derive the mechanical forces on the filament and the equations of motion.

\subsection{Mechanics}
\subsubsection{Passive elasticity}
We consider rods with an elastic free energy of the form
\begin{equation}
    F = \int \left[\frac{A}{2}\left(\kappa-\kappa_0\right)^2 + \Lambda\right]\mathrm{d}s\text{,}
\end{equation}
where $\kappa$ is the geodesic curvature, $\kappa_0$ the spontaneous geodesic curvature, $A$ is the bending modulus, $\Lambda$ is the tension in the rod and the arclength is given by $s$.

The stress tensor of such a rod is given by
\begin{equation}
    {\boldsymbol \sigma}^\text{el} =  \Lambda {\bf t} {\bf t} + A \partial_s\kappa {\bf t}{\bf n} - \frac{A}{2}\left(\kappa^2-\kappa_0^2\right){\bf t}{\bf t}\text{,}
\end{equation}
where, for example, ${\bf t} {\bf t}= {\bf t}\otimes {\bf t}$ is a rank-2 tensor. Taking the divergence with $\nabla = {\bf t}\partial_s$ gives the standard forces on a passive elastic rod
\begin{equation}
    {\bf f} = \nabla\cdot{\boldsymbol \sigma} = \partial_s \Lambda {\bf t} - \kappa \Lambda {\bf n} + A \left(\partial_s^2 \kappa + \frac{1}{2}\kappa\left(\kappa^2-\kappa_0^2\right)\right){\bf n}\text{.}
\end{equation}

Note that the asymmetric part of this stress corresponds to the classical Kirchoff bending moment tensor $\boldsymbol{\Phi}=A \kappa {\bf t} {\bf b}$ where ${\bf b}$ is the binormal, via the relation $\star\boldsymbol{\sigma}=\nabla\cdot\boldsymbol{\Phi}=$ where $\star$ is the Hodge dual.

\subsubsection{Odd stress}
We now include an active component of the elastic stress that is odd, in the sense that it breaks left-right symmetry. The simplest odd stress we can write down that breaks left-right symmetry must depend on even derivatives of the curvature (as if the stress depends on odd derivatives then it will give a force that is even in symmetry). This stress is of the form
\begin{equation}
    {\boldsymbol \sigma}^\text{odd}= \zeta \partial_s^{2}\kappa {\bf t}{\bf n}\text{,}
\end{equation}
which upon taking the divergence with $\nabla = {\bf t}\partial_s$ gives 
\begin{equation}
    {\bf f} = \nabla\cdot{\boldsymbol \sigma^{\mathrm{odd}}}  = \zeta \partial_s^{3} \kappa {\bf n} + \zeta \kappa \partial_s^{2} \kappa {\bf t}\text{.}
\end{equation}

This stress advects even derivatives of the curvature along the filament material frame. The lowest order term here can be seen as non-reciprocal analogue of the bending modulus. This corresponds to the following bending moment: $\boldsymbol{\Phi}=\zeta \nabla\kappa {\bf b}$.

\subsubsection{Viscous stress}
Adding the odd active stress preserves momentum but does not conserve energy, as such we need to include a dissipative mechanism to dissipate the energy injected by activity. This can be done in a momentum preserving manner with a bending viscosity.

This is given, in analogy with the elastic stress (to lowest order in curvature) and in agreement with other treatments in the literature~\cite{kodio_thesis,Buckmaster_Nachman_Ting_1975}, by
\begin{equation}
\boldsymbol{\sigma}^{\text{visc}} = \eta \partial_s \dot{\kappa} {\bf t} {\bf n} \text{,}
\end{equation}
which gives the following viscous force
\begin{equation}
    \nabla \cdot {\boldsymbol \sigma}^\text{visc} = \eta (\partial_s^2 \dot{\kappa}{\bf n} + \kappa \partial_s \dot{\kappa}{\bf t})\text{.}
\end{equation}

This corresponds to the bending moment $\boldsymbol{\Phi}=\eta \dot\kappa {\bf t} {\bf b}$.

\subsection{Dynamical equations}
The total stress tensor is given by ${\boldsymbol \sigma}={\boldsymbol \sigma}^\text{el}+ {\boldsymbol \sigma}^\text{visc}+{\boldsymbol \sigma}^\text{odd}$. Taking the divergence and balancing against substrate friction $\beta$ gives the force balance equation 
\begin{align}
    & \beta \dot{\bf x}=  \partial_s \Lambda {\bf t} - \kappa \Lambda {\bf n} + A \left(\partial_s^2 \kappa + \frac{1}{2}\kappa\left(\kappa^2-\kappa_0^2\right)\right){\bf n} +  \zeta \partial_s^{3} \kappa {\bf n} + \zeta \kappa \partial_s^{2} \kappa {\bf t}  + \eta (\partial_s^2 \dot{\kappa}{\bf n} + \kappa \partial_s \dot{\kappa}{\bf t})\text{,}
\end{align}
which must be solved along with the geometric and inextensibility conditions 
\begin{equation}
    {\bf t} = \partial_s {\bf x}\text{,} \quad {\bf n} = {\bf J }\cdot {\bf t}\text{,} \quad \dot{\bf x} = W {\bf t} + U {\bf n}\text{,} \quad \partial_s W + \kappa U = 0\text{.}
\end{equation}
We will rewrite these equations using the velocities $U, W$ as the primary variables:
\begin{align}
\beta W &= \partial_s \Lambda +\zeta \kappa \partial_s^{2} \kappa +\eta  \kappa\partial^2_s (W \kappa - \partial_s U),  \\
 \beta U &= - \Lambda  \kappa
 + A \left( \partial_s^2 \kappa + \frac{1}{2}\kappa\left(\kappa^2-\kappa_0^2\right)\right)\nonumber\\
 &\quad+ \zeta \partial^3_s\kappa+ \eta \partial_s^3 (W \kappa - \partial_s U), \\
 \partial_s W + \kappa U &=0. 
 \label{eq:ContinuumDynamics}
\end{align}
The explicit $\bf x$ dependence comes through updating positions using $W, U$:
\begin{equation}
\dot{\bf x} = W \partial_s {\bf x} + U {\bf J}\cdot  \partial_s {\bf x}
\end{equation}
with the curvature defined by
\begin{equation}
    \partial^2_s {\bf x} = - \kappa {\bf J}\cdot  \partial_s {\bf x}\text{.}
\end{equation}

\subsection{Linear stability analysis}
\label{sec:LSA}

First we consider the case of an infinite line parameterised by ${\bf x} = (x(s,t),h(s,t))$ where $h(s,t)\ll 1$. At zeroth order in height deformations the filament basis vectors are given by ${\bf t} = {\bf e}_x$ and ${\bf n}= {\bf e}_y$. The curvature is given by $\kappa = - h_{xx}$ and $U= \dot{h} + U_0$. For simplicity we will assume that the spontaneous curvature is zero ($\kappa_0=0$). The continuum dynamics~\eqref{eq:ContinuumDynamics} are solved exactly for $U=W=h=0$ with $\Lambda=\text{Const.}=\Lambda_0 + \delta \Lambda$. We will perturb around this base solution.

The continuity equation gives that $\partial_s W=0$ which leads to $\delta \Lambda = a x + b$ . The normal force balance equation decouples from the other equations giving 
\begin{equation}
     \beta \dot{h} =  \Lambda_0 h_{xx} - A h_{xxxx} - \zeta h_{xxxxx} - \eta\dot{h}_{xxxx}.
    \label{eq:LinearizedTheory}
\end{equation}
Taking the ansatz $h(x,t)=e^{(\lambda t + \mathrm{i} q x)}$ we obtain the dispersion
\begin{align}
    \lambda(q) = \frac{-1}{\beta+\eta q^4} \left(\Lambda q^2 + A q^4 + \mathrm{i}\zeta q^5  \right)\text{,}
    \label{eq:DispersionSI}
\end{align}
where $\lambda$ is the growth rate of the spatial Fourier mode with wavenumber $q$.~\eqref{eq:DispersionSI} appears as Eq. (3) of the Main Text. 

\section{The $\mathbb{Z}_2$ symmetric Bogdanov-Takens bifurcation}
\label{sec:Takens}
\begin{figure}[t]
    \centering
    \includegraphics[width=\linewidth]{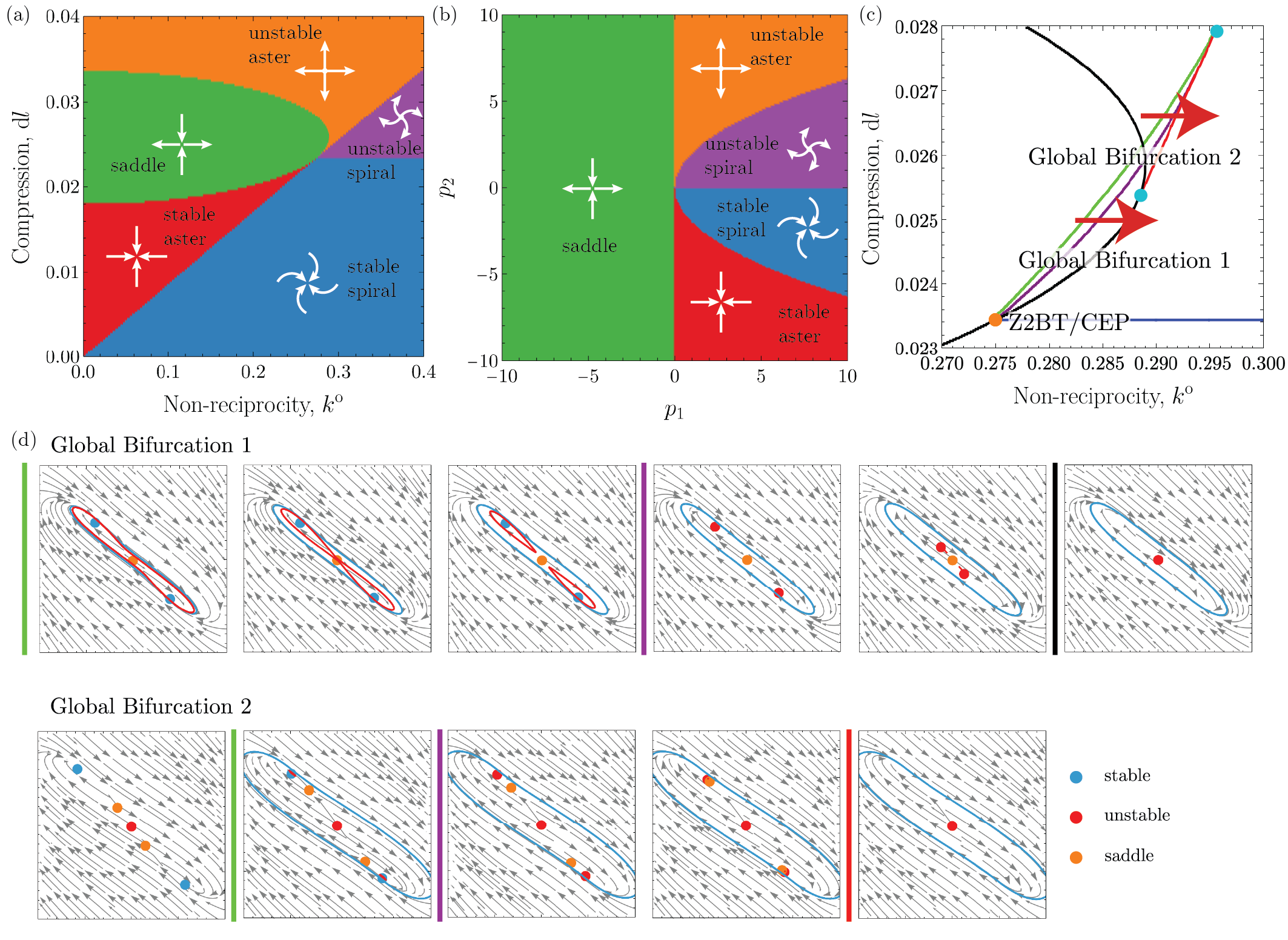}
    \caption{ {\bf Structure of the critical exceptional point}. (a) The linearized dynamics about the CEP in~\eqref{eq:EoMSI2}. Colours denote the structure of the fixed point at the origin, stable aster (red), saddle (green), unstable aster (organge), stable spiral (blue) and unstable spiral (purple), calculated according to the eigenvalues of $J$ at the origin. (b) Structure of the critical exceptional point in $\mathbb{Z}_2$ symmetric Bogdanov-Takens bifurcation. Colors are calculated as in (a). (c) Zoom in of the full system critical point, showing the region where the $\mathbb{Z}_2$ Bogdanov-Takens bifurcation occurs. Each line corresponds to a different type of bifurcation (for simplicity saddle connection bifurcations are omitted). Black: pitchfork bifurcation, blue: Hopf at origin, green: global saddle-node of two cycles, purple: symmetric Hopf bifurcations not at origin, red: saddle node bifurcation. (d) Phase plots showing the $(\theta_\mathrm{S},\theta_\mathrm{A})$ plane for the two paths shown in (c), that is varying $k^\mathrm{o}$ while leaving $\mathrm{d}l$ fixed. The colored lines between phase portraits correspond to the lines denoting bifurcations in panel (c).}
    \label{fig:EPSI}
\end{figure}
The CEP in the odd von Mises truss is a single point in parameter space, which occurs as part of a codimension-2 bifurcation scenario called a $\mathbb{Z}_2$ Symmetric Bogdanov-Takens bifurcation~\cite{golubitskyBurstingCoupledCell2005}. This bifurcation has a normal form
\begin{align}
    &y_1'=y_2\\
    &y_2'=-p_1y_1 +p_2 y_2 + a y_1^3 + b y_1^2y_2 +  c y_1 y_2^2 + d y_2^3
\end{align}with Jacobian given by
\begin{equation}
    {J}_{\text{BT}} = \left(
\begin{array}{cc}
 0 & 1 \\
 -p_1 & p_2 \\
\end{array}
\right)\text{.}
\label{eq:BTJ}
\end{equation}
By contrast, in the Main Text we derive the dynamics of our truss dynamics as: 
\begin{equation}
    \boldsymbol{\theta} = J\boldsymbol{\theta} + \sum_{j=0}^{3}
         {\bf A}_{m} \theta_\text{S}^m \theta_\text{A}^{3-m} \text{,}
\label{eq:EoMSI2}
\end{equation}
where $\boldsymbol{\theta}=(\theta_\mathrm{S},\theta_\mathrm{A})$. The coefficients $J$ and $\bf A$ read 
\begin{equation}
\begin{aligned}
    &J = \begin{pmatrix}
        \frac{dl k}{18}-1 & \left(\frac{5}{4}-3 {dl}\right)k^\mathrm{o}\\
        \left(-\frac{6 {dl} {}}{5}-{1}\right)k^\mathrm{o} & \frac{{dl} k}{30}-\frac{18 {dl}}{5}-1
    \end{pmatrix}\text{,}\\
    & {\bf A}_0 = \begin{pmatrix}
         \left(\frac{{1}}{8}-\frac{81 {}}{2 k}\right)k^\mathrm{o} \\
        -\frac{k}{120}-\frac{3}{4}
    \end{pmatrix}\text{,}\\
    & {\bf A}_1 = \begin{pmatrix}
        \frac{3}{2}-\frac{k}{72}\\
       \left( \frac{243 }{10 k}-\frac{9 }{10} \right) k^\mathrm{o}
    \end{pmatrix}\text{,}\\
    & {\bf A}_2 = \begin{pmatrix}
        \left(-\frac{81 {}}{2 k}-\frac{3 {}}{4}\right)k^\mathrm{o}\\
        \frac{9}{5}-\frac{k}{90}
    \end{pmatrix}\text{,}\\
    & {\bf A}_3 = \begin{pmatrix}
        -\frac{k}{54}\\
        \frac{k^\mathrm{o}}{5}
    \end{pmatrix}\text{.}
    \label{eq:Coefficients}
\end{aligned}
\end{equation}

In Fig.~\ref{fig:EPSI}(a, b) we compare the Jacobian of the truss~\eqref{eq:Coefficients} to that of the normal form~\eqref{eq:BTJ}, showing that they are related by a smooth deformation. For each case we compute the eigenvalues of the Jacobian for the fixed point at the origin and categorize into the following standard cases. Stable aster [two negative real eigenvalues, red in Fig.~\ref{fig:EPSI}(a, b)], saddle [one negative and one positive real eigenvalue, green in Fig.~\ref{fig:EPSI}(a, b)], unstable aster [two positive real eigenvalues, orange in Fig.~\ref{fig:EPSI}(a, b)], unstable spiral [two complex eigenvalues with positive real part, purple in Fig.~\ref{fig:EPSI}(a, b)], and stable spiral [two complex eigenvalues with negative real part, blue in Fig.~\ref{fig:EPSI}(a, b)].

The full details of the bifurcation structure in the truss are quite involved close to the CEP. Several global bifurcations occur in quick succession in the vicinity of the CEP, as shown in Fig.~\ref{fig:EPSI} (c, d). The black line in panel (c) denotes pitchfork bifurcations, the green a saddle node of two limit cycles (that is a pair of limit cycles form, one stable and one unstable), blue is the Hopf bifurcation at the origin, purple is the due Hopf bifurcations of the non-origin fixed points, red is a saddle node bifurcation. Phase portraits corresponding to two paths through this region (for different fixed $\mathrm{d}l$ varying $k^\mathrm{o}$) are show in Fig.~\ref{fig:EPSI}(d) with the colored lines between phase portraits corresponding to the bifurcations in Fig.~\ref{fig:EPSI}(c). The first path corresponds to the canonical global bifurcation behavior around a $\mathbb{Z}_2$ Bogdanov-Takens bifurcation, discussed in Ref.~\cite{golubitskyBurstingCoupledCell2005}, whereas the path labeled Global Bifurcation 2 is a higher order version of this with $5$ fixed points rather than $3$. Note that this narrow region between the green and purple lines corresponds to a region of stable buckled and snapping states. Eventually these lines merge with the fold (red line) to form the SNIC discussed in the main text.

Although this bifurcation behavior is very rich, it occurs in an exceptionally narrow region of the phase diagram, thus we do not discuss in detail in the main paper.

\section{Discrete simulations}\label{sec:DiscreteSimulations}
In this section, we perform additional discrete simulations of non-reciprocal beams which allow us to assess the robustness of assumptions made in our theoretical modelling. Our truss theory in the Main Text assumes small strain, and neglects inertia. Here we numerically validate these assumptions.

\subsection{Methodology}
\begin{figure*}[h!]
    \centering
    \includegraphics[width=\textwidth]{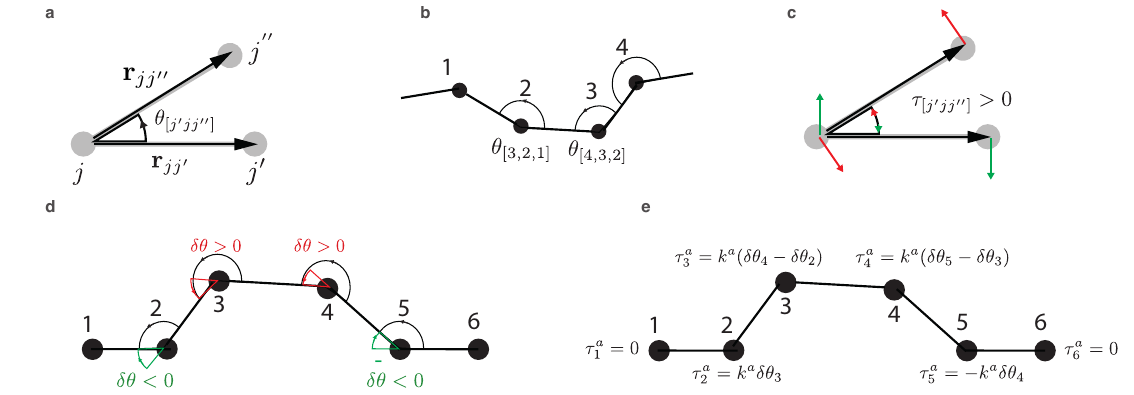}
    \caption{{\bf Schematic of a discrete odd chain}. Notation and variables associated with a discrete odd chain, including boundary conditions on angular tensions for an open chain. (a) Definition of an angle. (b) Example angles along a chain. (c) Example of an angular tension, with associated force dipoles (red/green). (d,e) In an open chain, we must prescribe boundary conditions for the active torques. Panel (d) shows the well-defined angles in an open chain made of 6 units, with the sense of angular deviations indicated (red/green). Panel (e) shows the associated active angular tensions generated at each vertex. At a chain end, we set all torques to zero. One unit in from the edge, we set undefined angular contributions to zero when evaluating the active torques.}
    \label{fig:DiscreteSimulationsSchematic}
\end{figure*}

A chain is described by a set of balls, labeled $j=1,2,...$ in a right-handed sense, and a set of springs connecting these balls. We consider two types of spring: longitudinal and torsional. Each longitudinal spring (labeled $\alpha=1,2,...$) identifies an ordered pair of ball indices $[j_\alpha,j_\alpha']$, and each torsional spring (labeled $\beta=1,2,...$) identifies an ordered triple of ball indices $[j_\beta',j_\beta,j_\beta'']$. A longitudinal spring $\alpha$ has a length change $\delta L_\alpha$ and linear tension $ T_\alpha$. A torsional spring $\beta$ has an angle change $\delta \theta_\beta $ and angular tension (or torque) $\tau_\beta$. Here, the angle subtended by this torsional bond is defined in a right-handed (counter-clockwise) sense, in other words from $j'$ to $j''$ about $j$. In practice, given the vectors $r_{jj'}$ and $r_{jj''}$, the angle $\theta_{j',j,j''} = \angle (r_{jj'}, r_{jj''})$ is unambiguously defined in a complex notation as
\begin{equation}
\theta_{[j',j,j'']} = \mathrm{Im} \mathrm{Log} \left( \frac{z_{jj''}}{z_{jj'}} \right) 
\end{equation}
which we implement via Numpy's angle function.

We consider a 1D chain of point particles of mass located at positions $\vb x_\alpha$, $\alpha =0 ,1 , \dots , N-1$.  We assume each bond has a rest length $l_0$, and we denote the angular tension in each vertex $\alpha$ by $\tau_\alpha$. In simulation, we consider extensible longitudinal bonds with a Hookean spring constant $k$, which we take to be large to approach the incompressible limit.

The equations of motion are the constructed via Newton's laws~\cite{veenstraAdaptiveLocomotionActive2025}, balancing inertia $m$ and sliding friction $\beta$ against elastic forces and viscous hinge dissipation:
\begin{equation}
\begin{aligned}
m \ddot {\vb x}_\alpha +\beta \dot{{\vb x}}_\alpha &= \epsilon \cdot \qty[ \frac{    \vb x_{\alpha +1} - \vb x_{\alpha } }{ \abs{\vb x_{\alpha +1} - \vb x_{\alpha }}^2 } ( \tau_\alpha (\underline{\vb x}) - \tau_{\alpha +1} (\underline{\vb x})  ) -  \frac{  \vb x_{\alpha -1} - \vb x_{\alpha } }{ \abs{\vb x_{\alpha -1} - \vb x_{\alpha }}^2 } ( \tau_\alpha (\underline{\vb x}) - \tau_{\alpha -1} (\underline{\vb x})  ) ]  \\
&+ k \qty[  \frac{ \vb x_{\alpha+1} - \vb x_{\alpha }  }{\abs{\vb x_{\alpha +1} - \vb x_\alpha } } ( \abs{\vb x_{\alpha +1} - \vb x_\alpha } -l_0  )+ \frac{ \vb x_{\alpha-1} - \vb x_{\alpha }  }{\abs{\vb x_{\alpha -1} - \vb x_\alpha } } ( \abs{\vb x_{\alpha -1} - \vb x_\alpha } -l_0 ) ] 
\label{eq:DiscreteEOM}
\end{aligned}
\end{equation}
where $\epsilon$ is the Levi-Civita symbol. The first terms in~\eqref{eq:DiscreteEOM} are the forces due to the angular tensions, while the second terms are the Hookean resistance to stretching. 
The functions $\tau_\alpha(\underline{\vb x})$ determine the vertex bending properties. For our non-reciprocal torsional elements, we model the vertices via
\begin{align}
    \tau_\alpha (\vb x) = B( \theta_\alpha - \theta^0) + k^\mathrm{o} ( \theta_{\alpha+1} - \theta_{\alpha-1}  ) + \Gamma {\dot \theta}_\alpha\label{eq:tau}.
\end{align}
Here $\theta^0$ is the rest angle (assumed to be identical for all vertices, typically taken as the initial angle), $B$ is the passive torsional elasticity, $\Gamma$ is the microscopic hinge dissipation, and $k^\mathrm{o}$ quantifies the non-reciprocity. 

For convenience, we list all microscopic and continuum parameters, alongside their dimension, in Table~\ref{tab:scales}. On dimensional grounds, and comparing to the linearized theory~\eqref{eq:LinearizedTheory}, we can make the following correspondence between the microscopic and continuum parameters:
\begin{align}
&\rho \sim \frac{m}{l_0}  \text{,}\\
&A \sim  B l_0 \text{,}\\
&\zeta \sim  k^\text{o} l_0^2\text{,} \\
&\eta \sim \Gamma l_0  \text{.}
\end{align}

\subsection{Non-dimensionalisation}
\begin{figure*}[t]
    \centering
    \includegraphics[width=\textwidth]{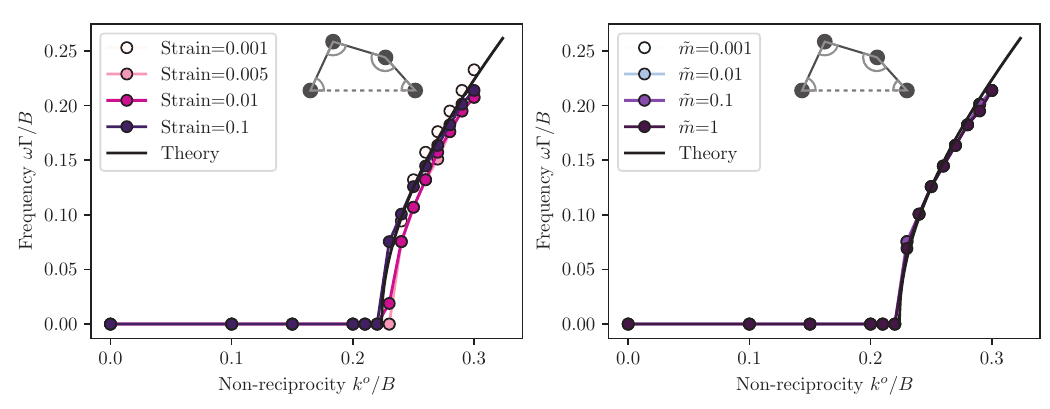}
    \caption{{\bf Effect of finite mass and compressive strain on the self-snapping instability.} Our theory~\eqref{eq:EoMSI2} is developed in the limit that $\tilde{m} \rightarrow 0 $ and of small compressive strain. However, our numerics show that our theory quantitatively captures the instability at $O(1)$ values of both parameters. In panel (a), $\tilde{m}=0.01$. In panel (b), strain is $0.1$. Throughout these simulations $\tilde{\beta}=0$, $\tilde{k}=1000$, with boundary compression implemented via stiff longitudinal springs with $\tilde{k}=5000$.}
    \label{fig:VaryingCompVaryingm}
\end{figure*}
We non-dimensionalise~\eqref{eq:DiscreteEOM} with a viscous timescale, and use the length of a single linkage as a characteristic length scale:
\begin{align}
&\mathrm{Viscous\ Timescale:} \quad \tau_v = \frac{\Gamma}{B} \\ 
&\mathrm{Lengthscale: } \quad l_0
\end{align}
These length and time scales are identical to those used in our microscopic theory of the von Mises truss, except that we non-dimensionalize by a single lattice spacing $l_0$ rather than the entire system size $l= N l_0$. The resulting dynamics is described by five dimensionless groups that mirror those used in our microscopic theory,
\begin{align}
&\tilde{k}^o = \frac{k^\mathrm{o}}{B}\text{,} \\
&\tilde{m} = \frac{m B l_0^2}{\Gamma^2}\text{,}\\
&\tilde{\beta} = \frac{\beta l_0}{\Gamma}\text{,} \\
&\tilde{k} = \frac{k l_0^2}{B}\text{,} \\
&N=l/l_0\text{,}
\end{align}
and reads
\begin{equation}
\begin{aligned}
\tilde{m} \ddot {\vb {x}}_\alpha +\tilde{\beta} \dot{{\vb x}}_\alpha &= \epsilon \cdot \qty[ \frac{    \vb x_{\alpha +1} - \vb x_{\alpha } }{ \abs{\vb x_{\alpha +1} - \vb x_{\alpha }}^2 } ( \tau_\alpha (\underline{\vb x}) - \tau_{\alpha +1} (\underline{\vb x})  ) -  \frac{  \vb x_{\alpha -1} - \vb x_{\alpha } }{ \abs{\vb x_{\alpha -1} - \vb x_{\alpha }}^2 } ( \tau_\alpha (\underline{\vb x}) - \tau_{\alpha -1} (\underline{\vb x})  ) ]  \nonumber \\
&+ \tilde{k} \qty[  \frac{ \vb x_{\alpha+1} - \vb x_{\alpha }  }{\abs{\vb x_{\alpha +1} - \vb x_\alpha } } ( \abs{\vb x_{\alpha +1} - \vb x_\alpha } -1  )+ \frac{ \vb x_{\alpha-1} - \vb x_{\alpha }  }{\abs{\vb x_{\alpha -1} - \vb x_\alpha } } ( \abs{\vb x_{\alpha -1} - \vb x_\alpha } -1 ) ],\\
\tau_\alpha (\vb x) &= ( \theta_\alpha - \theta^0) + \tilde{k}^o ( \theta_{\alpha+1} - \theta_{\alpha-1}  ) +  {\dot \theta}_\alpha.
\label{eq:Nondimensionaltau}
\end{aligned}
\end{equation}
In the Main Text, we focus on the limit $\tilde{\beta}=0$, and consider $\tilde{m}\rightarrow 0$. We refer to this limit as viscous. Because $\tilde{m}$ is dimensionally related to the $Q$ factor of a torsional oscillator as $\tilde{m}\sim Q^2$, this limit can be considered the $Q\rightarrow 0$ poor oscillator limit. In Ref.~\cite{veenstraAdaptiveLocomotionActive2025} an inertial instability of a viscously damped two-bar linkage is identified as occurring when 
\begin{equation}
\frac{k^\mathrm{o}}{\Gamma}\sqrt{\frac{m l_0^2}{B}} = \tilde{k}^o \sqrt{\tilde{m}} \geq 1.
\end{equation}
We avoid this instability by working in the $\tilde{m}\rightarrow 0$ limit. In Fig.~\ref{fig:VaryingCompVaryingm} we numerically explore the effects of varying the dimensionless mass $\tilde{m}$ and compressive strain on the self-snapping instability of the truss, in comparison to the viscous theory developed in the Main Text. We find that the instability is remarkably robust to both perturbations of our theory.

\section{Beam with substrate friction and follower forces}
\label{sec:Follower}
\begin{figure}[t]
    \centering
    \includegraphics{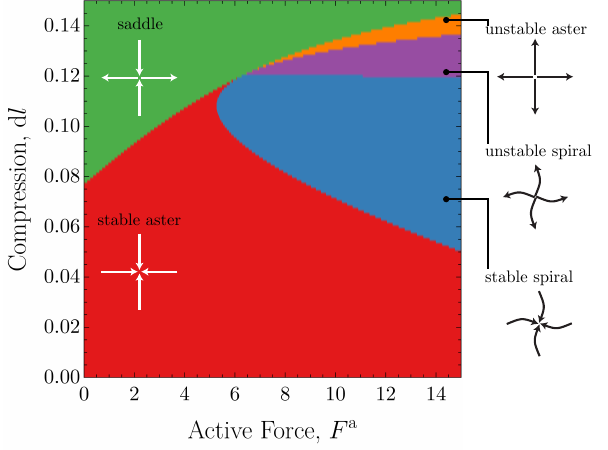}
    \caption{ {\bf Existence and structure of the critical exceptional point in a buckled follower force chain}. Classifying the Jacobian ~\eqref{eq:JFollowerForce} for the overdamped dynamics of a buckled follower force chain reveals a bifurcation scenario analogous to that of our non-reciprocal beam.} 
    \label{fig:FollowerForce}
\end{figure}

An important question is how generic is the bifurcation scenario described above. In order to address this question, we consider a beam with follower forces---also called tangential forces---which is extremely standard in the context of flagellar dynamics and active polymer literature~\cite{zhengSelfOscillationSynchronizationTransitions2023,sekimotoSymmetryBreakingInstabilities1995, decanioSpontaneousOscillationsElastic2017,active_poly_review2} with drag friction. In this case, the Rayleigh dissipation function is~\cite{zhengSelfOscillationSynchronizationTransitions2023}
\begin{align}
    \mathcal{R} = \frac{\eta}{2}\sum_{i=1}^{N} \dot x_\text{i}^2 + \dot y_\text{i}^2 + \sum_{i=2}^{N}-\vec{F^a_i}\cdot  \begin{pmatrix}
        \dot x_\text{i}\\
        \dot y_\text{i}
    \end{pmatrix}\text{,}
    \label{eq:Rayleigh}
\end{align}
where $\eta$ is the friction with the substrate, $-\vec{F^a_i}$ is the active force on node $i$ of magnitude $F^a$ and aligned with the bond vector connecting node $i$ to node $i-1$. $(x_\text{i}, y_\text{i})$ are the coordinates of node $i$, see Fig.~\ref{fig:DiscreteSimulationsSchematic} for a schematic. We now consider the case where nodes 1 and 4 are pinned, and express these in the same variables as defined for the odd truss in the Main Text:
\begin{align}
    \mathcal{R}&= \frac{\eta}{2}\left(w_L  \sec \theta_L \sin (2 \theta_L-\theta_R) \dot \theta_L\right)\nonumber\\&\qquad+F^a w_C^2 \left(\dot \theta_L-\dot \theta_L\right)^2 \sec ^2(\theta_L-\theta_R)+2 w_C w_L \dot \theta_L \sec (\theta_L) \left(\dot \theta_R-\dot \theta_L\right) \cos (2 \theta_L-\theta_R) \sec (\theta_L-\theta_R)+2 w_L^2 \dot \theta_L^2 \sec ^2(\theta_L),
    \end{align}
where we have used the same assumption as above $\dot w_C= \dot w_L=\dot w_R =0$. Using the same approach as described in the Main Text, we are able to derive the following equations of motion for the symmetric and antisymmetric angle $\theta_{S}$ and $\theta_{A}$:

\begin{align}
    \dot\theta_{S} &=  \frac{72 \delta l k (3 \delta l (F^a+48)-4 (F^a+42))+240 \delta l k^2-216 (F^a+20) k}{48 k}\theta_S\nonumber\\&\qquad
    -\frac{3}{4}  (\delta l (4 (3 F^a+k-105)-9 \delta l (F^a-60))+9 F^a-120)\theta_A+\frac{3 F^a (113 k-540)-10 k^2+432 (k+27)}{8 k}\theta_S\theta_A^2\nonumber
    \\&\qquad
    +\frac{\left(3 (37 F^a+250) k-324 (F^a+72)+4 k^2\right)}{4 k} \theta_A \theta_S^2+\frac{3 \left(3 (41 F^a-602) k-648 (F^a+1)+4 k^2\right) }{16 k}\theta_A^3\nonumber\\&\qquad+\frac{1}{6} (39 F^a-10 k+720) \theta_S^3 \nonumber\\
    \dot\theta_A &=  \left(-3 (\delta l-1) (3 \delta l (F^a+36)+F^a+24)-4 \delta l k\right)\theta_S+\frac{3}{2}  (-3 \delta l (3 \delta l-2) (F^a-40)+2 \delta l k+3 (F^a-20))\theta_A\nonumber\\&\qquad+\frac{\left(-9 (8 F^a+17) k+162 (F^a+6)+2 k^2\right) }{2 k}\theta_A^2\theta_S\nonumber\\&\qquad
    -\frac{(k (21 F^a+k+186)-3888) \theta_A \theta_S^2}{k}-\frac{3 (27 F^a (k-6)+(k-414) k+648) \theta_A^3}{4 k}-\frac{2}{3} (6 F^a-2 k+99) \theta_S^3.
\label{eq:EoMSI_follower}
\end{align}
We linearize these equations and find the Jacobian 
\begin{align}
J=
\left(
\begin{array}{cc}
 \frac{1}{2} (\delta l (9 \delta l (F^a+48)-12 (F^a+42)+10 k)-9 (F^a+20)) & \frac{3}{4} \left(9 \delta l^2 (F^a-60)-4 \delta l (3 F^a+k-105)-9 F^a+120\right) \\
 -3 (\delta l-1) (3 \delta l (F^a+36)+F^a+24)-4 \delta l k & \frac{3}{2} (-3 \delta l (3 \delta l-2) (F^a-40)+2 \delta l k+3 (F^a-20)) \\
\end{array}
\right).
\label{eq:JFollowerForce}
\end{align}
We use this Jacobian to determine the nature of the bifurcation scenario as a function of $F^a$ and $\delta l$ and find a qualitatively similar Bogdanov-Takens bifurcation scenario, all organized about a critical exceptional point (Fig.~\ref{fig:FollowerForce}). Hence, our results obtained in the case of non-reciprocal beam seem to pertain more generally to active filaments.

\section{List of Supplementary Videos}
\begin{itemize}
\item{Video 1: Non-reciprocal buckling.}
\item{Video 2: Breaking reciprocity in slender filaments creates one-way flexural waves.}
\item{Video 3: The odd von Mises truss is a minimal model of non-reciprocal buckling.}
\item{Video 4: Polyfunctions of non-reciprocal filaments.}
\item{Video 5: Crawling and stepping are robust to perturbations.}
\end{itemize}

\end{document}